\documentclass[oneside,twocolumn,aps,prl,preprintnumbers,superscriptaddress,amsmath,amssymb,10pt]{revtex4-2}

\usepackage{amsmath}
\usepackage{amsthm}
\usepackage{amssymb}
\usepackage{amsfonts}
\usepackage{bm}
\usepackage{braket}
\usepackage{cancel}
\usepackage{comment}
\usepackage{epstopdf}
\usepackage{float}
\usepackage{graphicx}
\usepackage{hyperref}
\usepackage{cleveref}
\usepackage{letltxmacro}
\usepackage{lipsum}
\usepackage{mathtools}
\usepackage{natbib}
\usepackage{physics}
\usepackage{subcaption}
\usepackage{tikz}
\usepackage{txfonts}
\usepackage{xcolor}

\begin{document}

\title{Nonclassical energy-change distribution as a witness of non-Markovian quantum dynamics}

\author{Marco Pezzutto}
\affiliation{Istituto Nazionale di Ottica del Consiglio Nazionale delle Ricerche (CNR-INO), Largo Enrico Fermi 6, I-50125 Firenze, Italy.}
\affiliation{Centro de Física e Engenharia de Materiais Avançados (CeFEMA), Laboratory of Physics for Materials and Emerging Technologies (LaPMET), and PQI -- Portuguese Quantum Institute, Portugal.}

\author{Anton Corr}
\affiliation{Centre for Quantum Materials and Technology, School of Mathematics and Physics, Queen's University Belfast, Belfast BT7 1NN, United Kingdom.}

\author{Gabriele De Chiara}
\affiliation{Física Teòrica: Informació i Fenòmens Quàntics, Departament de Física, Universitat Autònoma de Barcelona, 08193 Bellaterra, Spain.}
\affiliation{Centre for Quantum Materials and Technology, School of Mathematics and Physics, Queen's University Belfast, Belfast BT7 1NN, United Kingdom.}

\author{Salvatore Lorenzo}
\affiliation{Università degli Studi di Palermo, Dipartimento di Fisica e Chimica -- Emilio Segrè, via Archirafi 36, I-90123 Palermo, Italy.}

\author{Stefano Gherardini}
\affiliation{Istituto Nazionale di Ottica del Consiglio Nazionale delle Ricerche (CNR-INO), Largo Enrico Fermi 6, I-50125 Firenze, Italy.}
\affiliation{European Laboratory for Non-linear Spectroscopy, Università di Firenze, I-50019 Sesto Fiorentino, Italy.}

\begin{abstract}
We address the problem of identifying non-Markovian quantum time evolutions of an open quantum system by only performing measurements of the system's energy. We demonstrate that violations of CP-divisibility are always witnessed by non-positive values of the energy-change Kirkwood-Dirac quasiprobability distribution associated with the system's Hamiltonian, evaluated at consecutive times. The link between non CP-divisibility and non-positivity of the system's energy-change distribution is stronger when the system-environment interactions are energy-preserving. The witness works whenever anomalous energy fluxes, due to non-Markovianity, are realized. Anomalous fluxes are also detected by the non-Markovianity measure built over the quantum mutual information between the states of the open system and of a quantum correlated reference. 
\end{abstract}

\maketitle

In many-body physics and thermodynamics, characterizing local properties of the quantum system under scrutiny can give access to collective and dynamical behaviors associated to observables such as energy, magnetization, population imbalance, etc. Moreover, within currently platforms such as quantum simulators~\cite{AltmanPRXQ2021} and annealers~\cite{Rajak2023}, measuring properties that also incorporate the interactions between the system and the environment still remains operationally inaccessible.

In this Letter, we focus on detecting memory effects in the evolution of an open quantum system using the information obtained by performing measurements of the system's energy. In particular, we consider a composite non-Markovian collision model~\cite{ciccarello2022a,cusimano2022} in which the system interacts with an auxiliary memory that in turn collides sequentially with fresh environment particles~\cite{LorenzoPRA2016,lorenzo2015b,lorenzo2017g}. Considering exchange interactions --- prevalent in spin, atomic, optical and QED systems~\cite{Heisenberg28,Baxter1982,RaimondRMP2001,Bernardes2015,MaffeiPRR2021} --- and qubits, the reduced collision map turns out to be a phase covariant map~\cite{filippov2020,Smirne2016,siudzinska2023,Holevo1993}, which allows us for a rigorous treatment of non-Markovianity~\cite{rivas_2014,shrikant2023} in the presence of system-environment correlations.

This framework is not merely phenomenological: in quantum optical settings it arises naturally when the memory is identified with a localized cavity, resonator, or pseudo-mode degrees of freedom, while the environment particles represent successive time bins of the traveling field~\cite{Ciccarello2017}. 
This model is also consistent with realistic repeated-interaction platforms: e.g., in the one-atom maser, individual Rydberg atoms interact sequentially with a high-quality cavity mode~\cite{Meschede1985,RaimondRMP2001}, while in superconducting structured-reservoir experiments non-Markovian feedback has been observed on directly accessible experimental timescales~\cite{Ferreira2021}. 
More generally, recent rigorous results show that collision models can be derived from microscopic Hamiltonians, becoming numerically exact when the collision time step is chosen sufficiently small to resolve the relevant timescales of the environment~\cite{CattaneoPRL2021,Lacroix2025}.

We identify analytical conditions under which the loss of positivity of the Kirkwood-Dirac Quasiprobability (KDQ) distribution for the system's energy changes~\cite{LevyPRX2020,LostaglioQuantum2023,hernandez2022experimental,SantiniPRB2023,GherardiniTutorial,HernandezNpjQI2024,ArvidssonShukur2024review,Pezzutto2025non-positive,Yoshimura2025,Perciavalle2025,DonelliPRA2026} witnesses non-Markovianity of the system's dynamics. Here, non-Markovianity is quantified in two ways: in terms of the non-Completely Positive (CP) divisibility of the system's dynamics due to entanglement generation [Rivas-Huelga-Plenio (RHP) measure]~\cite{RHP-non-Markovianity}, and in terms of the quantum mutual information (QMI) between the system and a correlated reference system [Luo-Fu-Song (LFS) measure]~\cite{LuoPRA2012}.

We associate the loss of positivity in the system's energy-change distribution with the presence of {\it anomalous energy fluxes} from/to the system~\cite{LevyPRX2020,solinas2022,GherardiniTutorial,ComarPRXQuantum2025,Mallik2025}. Here, the anomaly refers to thermalization processes involving cooling (heating), whereby the system emits (absorbs) energy to (from) the bath/environment/reservoir in a non-monotonic manner, thus alternating emissions and absorptions. This is the signature of reversing fluxes~\cite{HuangPRL2026}, occurring locally in time due to non-Markovian temporal correlations. In this analysis, the system is initialized in a thermal state, so that the presence of quantum coherence in the initial state does not play a role.

Time-local reversing fluxes are linked with the non-monotonicity of QMI for the bipartite state of the open system and an auxiliary one. This proves that the positivity loss of the KDQ distribution, non-CP-divisibility and the onset of anomalous energy fluxes originate from the system's tendency to correlate with the environment during the dynamics.

Our results explain the nonclassical behaviors of collision models exhibiting non-Markovianity~\cite{LiPRA2024,Lalita2025}. They also provide the analytical conditions under which one can witness the presence of memory effects in an open system's dynamics solely via local measurements, while studying the corresponding energy-change statistics.

{\bf Collision models \& phase covariant maps---} 
Consider the quantum system $S$ interacting with an environment composed of many particles ${\{}A^{(n)}{\}}$ via an intermediate system, denoted as the memory $M$~\cite{LorenzoPRA2016,lorenzo2015b}. The memory collides sequentially with different environmental particles, each of which is discarded after the interaction. Hence, before each collision with the memory, the environment particle is always in the same state $\rho_{A}$. $S, M, A^{(n)}$ are taken as spin-1/2 particles, governed by the local Hamiltonians $H_{k} = (\hbar \omega_k / 2)\sigma^z_k$ with $k{=}S, M, A$ and $\sigma^{\alpha}_k$ is the Pauli matrix along the direction $\alpha{=}x,y,z$. The sum of local Hamiltonians is $H_0 {=} H_S+H_M+H_A$.
The system's dynamics occurs thanks to pairwise interactions between $S$ and $M$, and between $M$ and $A$, which are turned on and off cyclically. The interaction Hamiltonians are chosen as the Heisenberg interaction terms $H_{kj} {=} (\hbar g_{kj}/2) \left( \sigma_k^x \otimes \sigma^x_j {+} \sigma_k^y \otimes \sigma^y_j {+} \sigma_k^z \otimes \sigma^z_j \right)$ where $k\!j = S\!M, M\!A$ and $g_{kj}$ denote the interaction strengths.

The dynamics are described in discrete time steps $t=n\tau$, with $n=0,1,\dots$; see also the Appendix. At the $n^{\text{th}}$ step, the system $S$ first interacts with the memory $M$, and the memory then interacts with the environmental particle $A^{(n)}$. These two interactions are generated by the unitary operators
\begin{align}
U_{SM}(\tau_1) &= \exp \left( -\frac{ i }{\hbar} ( H_0 + H_{SM}\otimes \mathbb{I}_{A} ) \tau_1 \right) ,\label{eq:U_SM}
\\
U_{MA}^{(n)}(\tau_2) &=  \exp \left( -\frac{ i }{\hbar} ( H_0 + \mathbb{I}_{S}\otimes H_{MA}  ) \tau_2 \right)_.\label{eq:U_MA}    
\end{align}
In Eq.~\eqref{eq:U_MA}, the superscript $(n)$ in $U_{MA}^{(n)}$ indicates that the memory interacts with the $n^{\text{th}}$ environmental particle, $A^{(n)}$, and the duration of one full time step is $\tau{=}\tau_1+\tau_2$. When no ambiguity arises, we will omit the explicit dependence of the unitary operators on $\tau_1$ and $\tau_2$. The Heisenberg interactions generated by the unitary operators in Eqs.~\eqref{eq:U_SM} and \eqref{eq:U_MA} implement a \emph{partial coherent swap}. Specifically, for the $k$-th and $j$-th subsystems ($k,j = S,M,A^{(n)}$), the Heisenberg interaction $H_{kj}$ generates a perfect swap when the interaction time $\tilde{\tau}$ is chosen such that $g_{kj}\tilde{\tau}=\pi/2$. In this case, $U_{kj}(\tilde{\tau}) \left( \rho_k \otimes \rho_j \right)U_{kj}(\tilde{\tau})^\dagger = \rho_j \otimes \rho_k$. For other values of $g_{kj}\tilde{\tau}$, the same interaction realizes a coherent partial swap, which in general builds quantum correlations between the two subsystems.
The fully Markovian limit is recovered when $U_{MA}$ implements a perfect swap between the memory $M$ and a fresh environmental particle $A$. In this case, the memory state is completely replaced by that of the incoming environmental particle. Consequently, in the following step, the interaction between the system and the memory is equivalent to a direct interaction between the system and a fresh environmental particle.

Using the matrix-element representation for operators on $S$, namely $\ket{\rho_S}\!\rangle = (\rho_{00},\rho_{01},\rho_{10},\rho_{11})^T$, one can show that the populations $\rho_{00}$ and $\rho_{11}$ couple only to each other. By contrast, coherences $\rho_{01}$ and $\rho_{10}$ evolve independently and remain decoupled from populations~\cite{filippov2020}; see also the Appendix. Accordingly, the most general Hermiticity- and trace-preserving qubit map can be written as the \emph{phase covariant map}
\begin{equation}
\label{eq:ansatz}
\Lambda=
\begin{pmatrix}
a & 0 & 0 & b\\
0 & c & 0 & 0\\
0 & 0 & c^* & 0\\
1-a & 0 & 0 & 1-b
\end{pmatrix}.
\end{equation}

{\bf Quantum statistics of system's energy changes---} 
The stochastic system's energy changes $u_S$ between collisions are given by all possible combinations of differences of the eigenvalues of the bare Hamiltonian operator $H_S$. We thus consider the spectral decomposition of $H_S$: $H_S = \sum_{\ell}E^{S}_{\ell}\Pi^{S}_{\ell}$. Focusing on single collisions, $u_S$ are operationally evaluated immediately before the interaction is activated ($\ell=\ell_{\rm in}$), and immediately after it is switched off ($\ell=\ell_{\rm fin}$). Thus, we consider the energy exchanges between iterations $n-1$ and $n$. Since $H_S$ is time-independent, one has that $u_S (\ell_{\rm in},\ell_{\rm fin}) = E^{S}_{\ell_{\rm fin}} - E^{S}_{\ell_{\rm in}}$. For the $n^{\text{th}}$ collision, the distribution $P_n(u_S)$ of $u_S$ is defined as
\begin{equation}
\label{eq:distribution}
    P_n(u_S) = \sum_{\ell_{\rm in},\ell_{\rm fin}} \mathfrak{q}_n \big( u_S( \ell_{\rm in}, \ell_{\rm fin} ) \big) \, \delta\big( u_S - u_S( \ell_{\rm in}, \ell_{\rm fin} ) \big),
\end{equation}
where $\delta(\cdot)$ is the Dirac delta function, and $\Big\{ \mathfrak{q}_n\left( u_S( \ell_{\rm in}, \ell_{\rm fin} ) \right) \Big\}$ are KDQs~\cite{LostaglioQuantum2023,GherardiniTutorial,ArvidssonShukur2024review}. They describe the two-time probability of sequentially recording $u_S$, without erasing quantum coherences in the first measured state.

The quasiprobabilities $\mathfrak{q}\left( u_S (\ell_{\rm in},\ell_{\rm fin}) \right)$ read as
\begin{equation}\label{eq:QP_uS_map}
    \mathfrak{q}\left( u_S (\ell_{\rm in},\ell_{\rm fin}) \right) {=} {\rm Tr}\left[  \Pi^{S}_{\ell_{\rm fin}}   \Lambda[\Pi^{S}_{\ell_{\rm in}}  
    \rho_{S} ] \right] 
    {=} {\rm Tr}\left[ \Lambda^{\dagger}[ \Pi^{S}_{\ell_{\rm fin}}]   \Pi^{S}_{\ell_{\rm in}}  
    \rho_{S}  \right],    
\end{equation}
where $\Lambda^{\dagger}$ denotes the dual map of the quantum channel $\Lambda$, and $\rho_{S}$ is the system state just before the collision. Note that $\Lambda^{\dagger}$ acts on operators (instead of states) according to the Heisenberg picture. The channel $\Lambda$, reconstructed through tomography as described in the Appendix, contains all the information about any non-Markovian effect entailed by the interaction of the system with the environment particles, mediated by the memory. In particular, we denote by $\{ \Lambda_n \}$ the family of maps that evolve the system from the initial state up to the state after $n$ collisions, and by $\Lambda_n^{-1}$ the inverse map of $\Lambda_n$. Moreover, the quasiprobabilities \eqref{eq:QP_uS_map} for energy changes from the $(n-1)^{\text{th}}$ to the $n^{\text{th}}$ step take the expression $\mathfrak{q}_n\left( u_S (\ell_{\rm in},\ell_{\rm fin}) \right) = {\rm Tr}\left[  \Pi^{S}_{\ell_{\rm fin}} \Lambda_{n,n-1}\left[ \Pi^{S}_{\ell_{\rm in}} \rho_S^{(n-1)} \right] \right]$, where $\Lambda_{n,n-1} = \Lambda_n \circ \Lambda^{-1}_{n-1}$ with $\circ$ denoting the map composition.

When the state $\rho_S^{(n)}$ of the measured system does not commute with the measurement observable $H_S$, the KDQs $\mathfrak{q}_n$ could lose positivity, becoming complex with negative real parts. This relevant feature can be employed to detect whether quantum effects impact the dynamics of the system, through the \emph{non-positivity functional}~\cite{LostaglioQuantum2023,GherardiniTutorial}
\begin{equation}
\label{eq:nonclassical}
\mathcal{N_{\rm q}}\Big[ P_n( u_S ) \Big] \equiv -1 + \sum_{\ell_{\rm in},\ell_{\rm fin}} \Big\vert \mathfrak{q}_n \Big( u_S(\ell_{\rm in},\ell_{\rm fin})  \Big) \Big\vert.
\end{equation}
In the following, we will show how $\mathcal{N_{\rm q}}\Big[ P_n( u_S ) \Big]$ can detect the emergence of non-Markovianity in the system dynamics.

{\bf Non-positivity as a witness of non-Markovianity---}
Let us assume the system to be in the state $\rho_S^{(n-1)}$ after the $(n-1)^{\text{th}}$ collision. For phase covariant maps [Eq.~\eqref{eq:ansatz}], the direct calculation of quasiprobabilities \eqref{eq:QP_uS_map}, with $E^S_0=-\hbar\omega_S/2$ and $E^S_1=\hbar\omega_S/2$, leads to 
\begin{equation}
\label{eq:KD_explicit}
\begin{aligned}
\mathfrak q_n\left( E^S_0 {-} E^S_0\right)=a_n\,p_0^{(n-1)},\,\,\,\,
\mathfrak q_n\left( E^S_1 {-} E^S_0\right)=(1-a_n)\,p_0^{(n-1)},\\
\mathfrak q_n\left( E^S_0 {-} E^S_1\right)=b_n\,p_1^{(n-1)},\,\,\,\,
\mathfrak q_n\left( E^S_1 {-} E^S_1\right)=(1-b_n)\,p_1^{(n-1)}.
\end{aligned}
\end{equation}
In Eq.~\eqref{eq:KD_explicit}, $a_n$ and $b_n$ are the entries of the single-collision map $\Lambda_{n,n-1}$ that act on the diagonal elements, $\rho_{00}^{(n-1)}$ and $\rho_{11}^{(n-1)}$, of the state $\rho_S^{(n-1)}$; see Eq.~\eqref{eq:ansatz}. Instead, $p_0^{(n-1)}, p_1^{(n-1)}$ are the probabilities to measure the energies $E^S_0, E^S_1$ after the $(n-1)^{\text{th}}$ collision. These probabilities are given by the Born rule, and thus are real, positive numbers belonging to $[0,1]$. They correspond to the diagonal elements of the pre-collision state. As a result, the KDQs $\mathfrak{q}_n(u_S)$ are insensitive to the parameter $c_n$ linked with the quantum coherence $\rho_{01}^{(n-1)}$.

Using \eqref{eq:KD_explicit}, the non-positive functional \eqref{eq:nonclassical} becomes
\begin{equation}\label{eq:Nq_explicit}
\mathcal{N_{\rm q}}\Big[ P_n( u_S ) \Big] 
=-1 + p_0^{(n-1)}\Big(|a_n|+|1-a_n|\Big) + p_1^{(n-1)}\Big(|b_n|+|1-b_n|\Big). 
\end{equation}
If $a_n<0$ or $a_n>1$, then either $\mathfrak q_n( E^S_0 - E^S_0)$ or $\mathfrak q_n( E^S_1 - E^S_0)$ is negative. Similarly, if $b_n<0$ or $b_n>1$, then either $\mathfrak q_n( E^S_0 - E^S_1)$ or $\mathfrak q_n( E^S_1 - E^S_1)$ is negative. Therefore, a necessary and sufficient condition for the non-positivity of the KDQ distribution $P_n(u_S)$ is:
\begin{equation}
\label{eq:KD_nonpositivity_conditions}
\mathcal{N_{\rm q}}\Big[ P_n( u_S ) \Big]  > 0
\;\; \Longleftrightarrow \;\;
\text{either } a_n\notin[0,1]\ \text{or}\ b_n\notin[0,1]\,.
\end{equation}

To determine the complete positivity of the map $\Lambda_{n,n-1}$, we compute the corresponding Choi matrix: 
\begin{equation}
\label{eq:Choi}
J_{\Lambda_{n,n-1}} =
\sum_{i,j=0}^1 \ketbra{i}{j}\otimes \Lambda_{n,n-1}\Big[ \dyad{i}{j} \Big]
= \begin{pmatrix}
a_n & 0 & 0 & c_n\\
0 & 1{-}a_n & 0 & 0\\
0 & 0 & b_n & 0\\
c_n^* & 0 & 0 & 1{-}b_n
\end{pmatrix}.
\end{equation}
The map $\Lambda_{n,n-1}$ is completely positive if and only if $J_{\Lambda_{n,n-1}} \ge 0$. This is equivalent to
\begin{equation}
\label{eq:CP_conditions}
0\le a_n\le 1,\qquad
0\le b_n\le 1,\qquad
|c_n|^2\le a_n(1-b_n).
\end{equation}
The first two conditions concern the sector of the map operating on diagonal elements of $\rho_S$, whereas the last condition concerns the sector of the map that acts on quantum coherence terms. Therefore, a breakdown of complete positivity may originate either from diagonal or off-diagonal elements of $\rho_S$ after the repeated application of the dynamical map.

In conclusion, comparing relations \eqref{eq:KD_nonpositivity_conditions} and \eqref{eq:CP_conditions}, we can conclude that if $\mathcal N_{\mathrm q}[P_n(u_S)] > 0$, then $J_{\Lambda_{n,n-1}}$ is not positive and hence $\Lambda_{n,n-1}$ is not completely positive:
\begin{equation}
\label{eq:KDvsCP}
\mathcal N_{\rm q}>0 \;\Longrightarrow\; \Lambda_{n,n-1} \text{ is not CP} \,\,\, \text{(for any $n\geq 0$)}.
\end{equation}
Eq.~\eqref{eq:KDvsCP} is the main result of the Letter. Violations of CP-divisibility are a hallmark of non-Markovian quantum dynamics, and are quantified by the RHP measure~\cite{RHP-non-Markovianity}: $\mathcal{I_{\rm RHP}} = \sum_{n}g_{n}$, where $g_n = \Vert (\Lambda_{n,n-1} \otimes \mathbb{I}) \big[ \ketbra{\Psi}{\Psi} \big] \Vert_1 - 1$ with $\ket{\Psi}$ denoting a maximally entangled state. In the Appendix we report the definition of the RHP measure for discrete-time processes.

\begin{figure}[t]
    \centering
    \includegraphics[width=0.85\columnwidth]{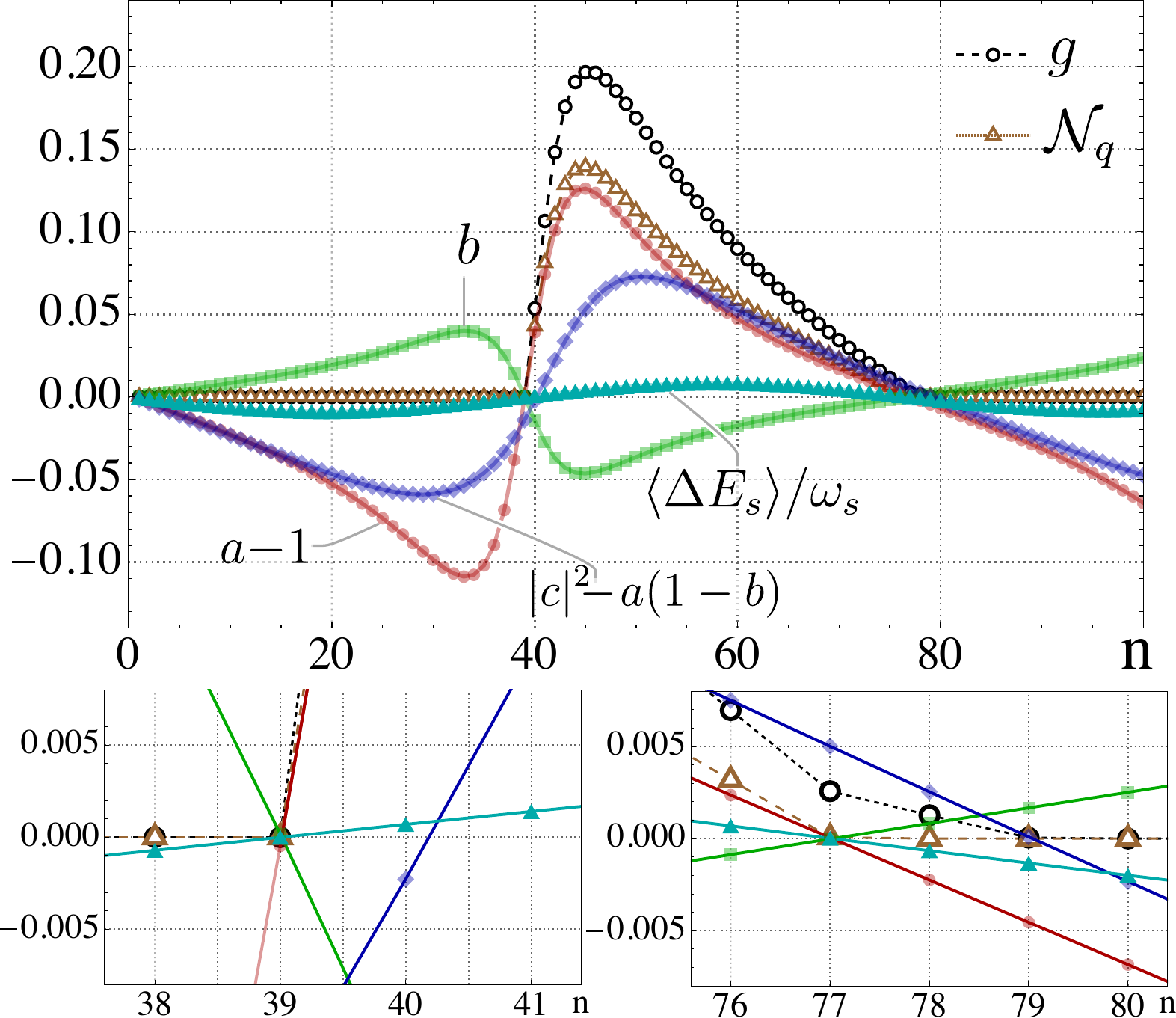}
    \caption{
    Energy-preserving interactions. {\it Top panel}: Comparison between non-CP-divisibility identified by $g_n$ (black circles), and the non-positivity functional $\mathcal N_{\mathrm q}[P_n]$ of the system's energy-change distribution (brown triangles). 
    {\it Bottom panels}: zoomed-in view of the top panel where $\mathcal N_{\mathrm q}[P_n]$ becomes nonzero (on the left) and returns to zero (on the right). In all panels we report the conditions \eqref{eq:CP_conditions} (red dots, green squares and blue rhombs, respectively), as well as the average system's energy-change $\langle\Delta E_S\rangle$ normalized by $\omega_S$ (teal triangles). 
    Parameters: $\omega_S {=} \omega_M {=} \omega_A {=} \omega {=} 1$ (resonant interactions); $g_{SM} {=} g_{MA} {=} 0.2 \omega$, and $\tau_1 {=} \tau_2 {=} 0.2 /\omega$ (well within the weak coupling limit); $\beta{=}1$. In this figure, as in the others throughout the Letter, the frequencies and temperatures are expressed in natural units such that the reduced Planck constant and Boltzmann constant are $\hbar=1, k_B=1$, respectively.
    }
    \label{fig:nonMK-nonpos}
\end{figure}

{\bf Physical interpretation via anomalous energy fluxes---}In Fig.~\ref{fig:nonMK-nonpos}, we show the capability of the non-positive functional $\mathcal N_{\mathrm q}[P_n(u_S)]$ to detect violations of CP-divisibility, for a cooling process modeled by a memory-mediated collision model. The system's initial state is the completely mixed state $\mathbb{I}/2$, while the memory and environment particles are initialized in the thermal state at inverse temperature $\beta$. Then, resonant Heisenberg interactions are considered, i.e., $\omega_k=\omega$ $\forall k$.

From Fig.~\ref{fig:nonMK-nonpos}, we observe that the non-CP-divisibility of the system's dynamics is mainly due to violations of the conditions $0\le a_n\le 1$ and $0\le b_n\le 1$. This entails that $\mathcal N_{\mathrm q}[P_n(u_S)]$ and $g_n$ are different from zero within the same range of collisions: they start to grow together at the $39$th collision (left bottom panel), and come back to zero at the $77$th and $79$th collisions, respectively (right bottom panel). In Fig.~\ref{fig:nonMK-nonpos}, we also plot the system's average energy-change  
\begin{eqnarray}\label{eq: DeltaES}
    \langle\Delta E_S\rangle 
    &\equiv \Tr\left[ H_S \left(\Lambda_{n,n-1}\left[ \rho_S^{(n-1)} \right]{-}\rho_S^{(n-1)}\right)\right]=\nonumber\\
    &=\omega_S \left[ (a_n-1)p_0^{(n-1)} + b_n \, p_1^{(n-1)} \right].
\end{eqnarray}
The sign change of $\langle\Delta E_S\rangle$ reveals the presence of an anomalous energy flux between the system and the environment. Anomalous energy flux occurs exactly whenever $\mathcal N_{\mathrm q} \neq 0$, and in this context is the signature of a non-Markovian process.

{\bf Non-monotonicity of quantum mutual information---}
Following \cite{PhysRevA.86.044101}, consider an auxiliary system $L$ that is initially correlated with the system $S$. The total correlations of the bipartite state $\rho_{LS}$ are quantified by the QMI $I(\rho_{LS}) = S(\rho_L) + S(\rho_S) - S(\rho_{LS})$, where $S(\rho) = -\rm{Tr}[\rho \log_2(\rho)]$ is the von-Neumann entropy. If the map $\Lambda_n$ is Markovian, then $\Delta I_{LS}^{(n)} \equiv I(\rho_{LS}^{(n+1)})-I(\rho_{LS}^{(n)})\leq 0$ for any $n$, due to the monotonicity of QMI under local operations. Any violation indicates non-Markovianity and is quantified by $\mathcal{I}_{\text{LFS}} = \sum_{k\,\text{s.t.}\,\Delta I_{LS}^{(k)}>0}\Delta I_{LS}^{(k)}$, with $S, L$ initialized in an arbitrary maximally correlated pure state $\rho_{LS}^{(0)}$; see also the Appendix. 

\begin{figure}[t]
    \centering
    \includegraphics[width=0.85\columnwidth]{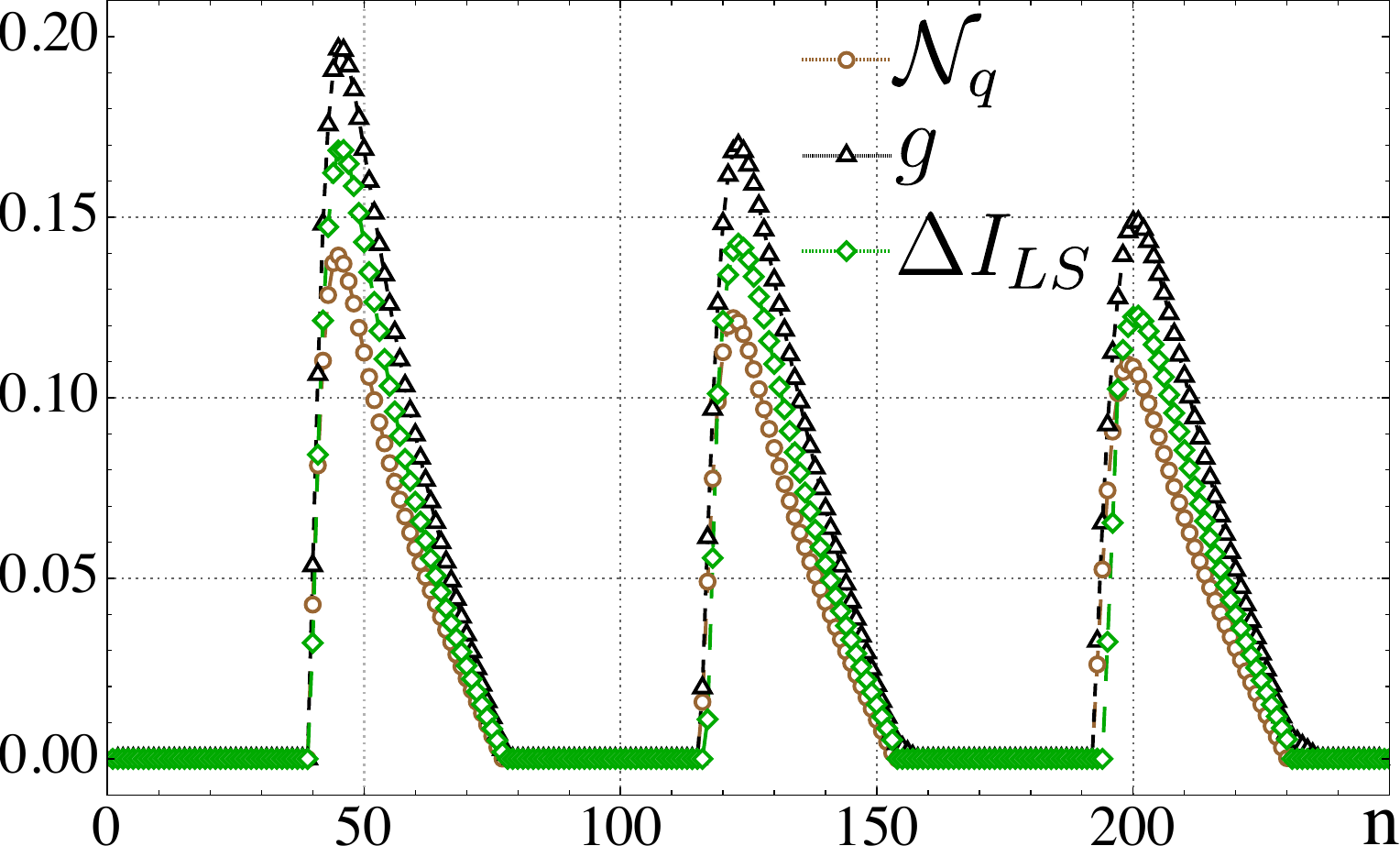}
    \caption{
    Energy-preserving interactions. Comparison of $\Delta I_{LS}^{(n)}$ (green rhombs), $g_n$ (black triangles) and $\mathcal N_{\mathrm q}[P_n]$ (brown circles), as a function of collision number. The initial states of $S, M, A$ and model's parameters are the same as in Fig.~\ref{fig:nonMK-nonpos}.
    }
    \label{fig:comparison_LFS}
\end{figure}    

The onset of anomalous energy fluxes means not only violation of CP-divisibility, but even the non-monotonic decrease of the QMI $I(\rho_{LS})$ under local operations. We show this in Fig.~\ref{fig:comparison_LFS}, where we compare $\Delta I_{LS}^{(n)}$ with $g_n$ and $\mathcal N_{\mathrm q}[P_n(u_S)]$, in a regime with energy-preserving interactions. All these quantities almost exactly overlap. 

\begin{figure}[t]
    \centering
    \includegraphics[width=0.875\columnwidth]{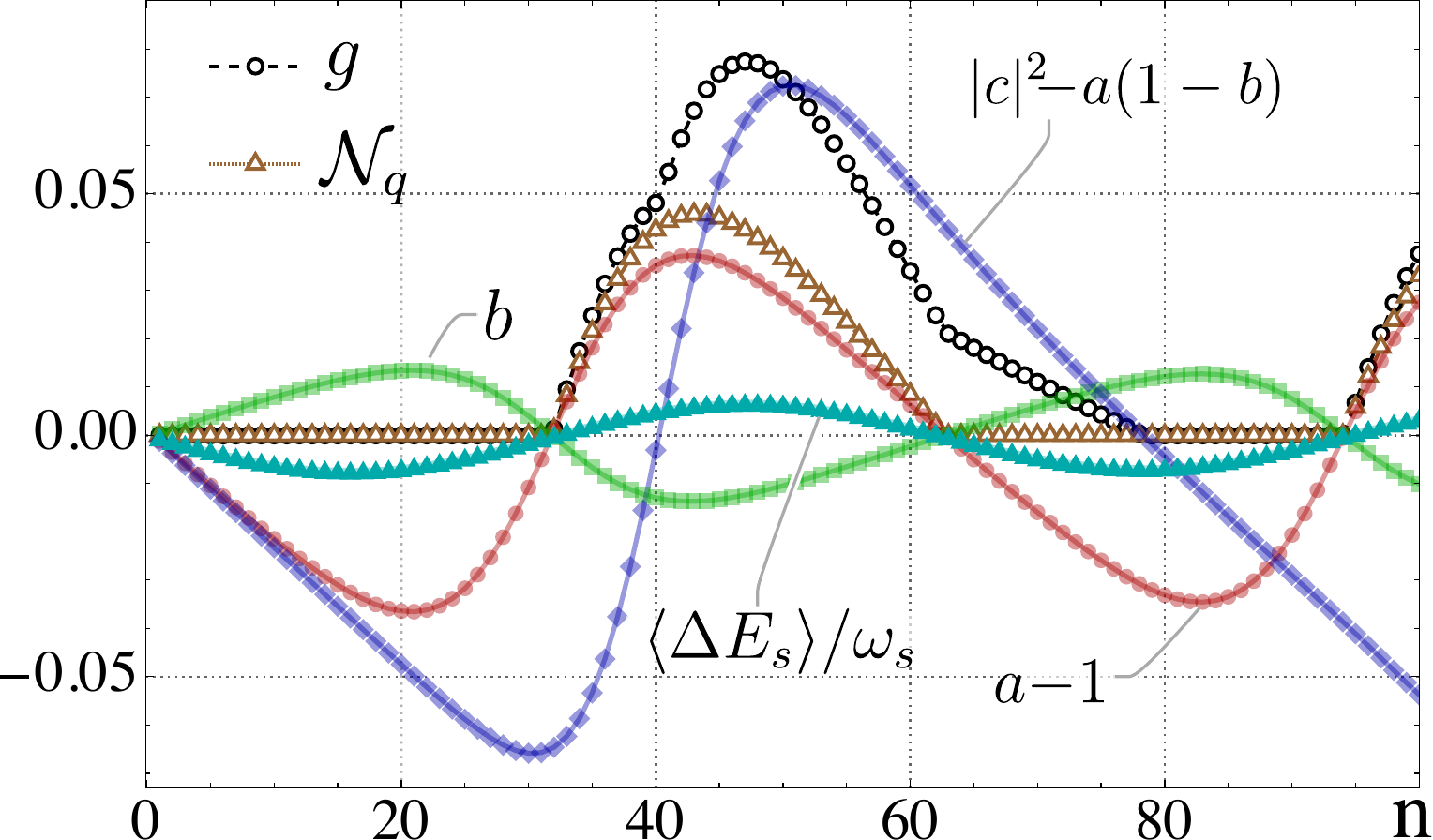}
    \\
    \vspace{0.3cm}
    \includegraphics[width=0.875\columnwidth]{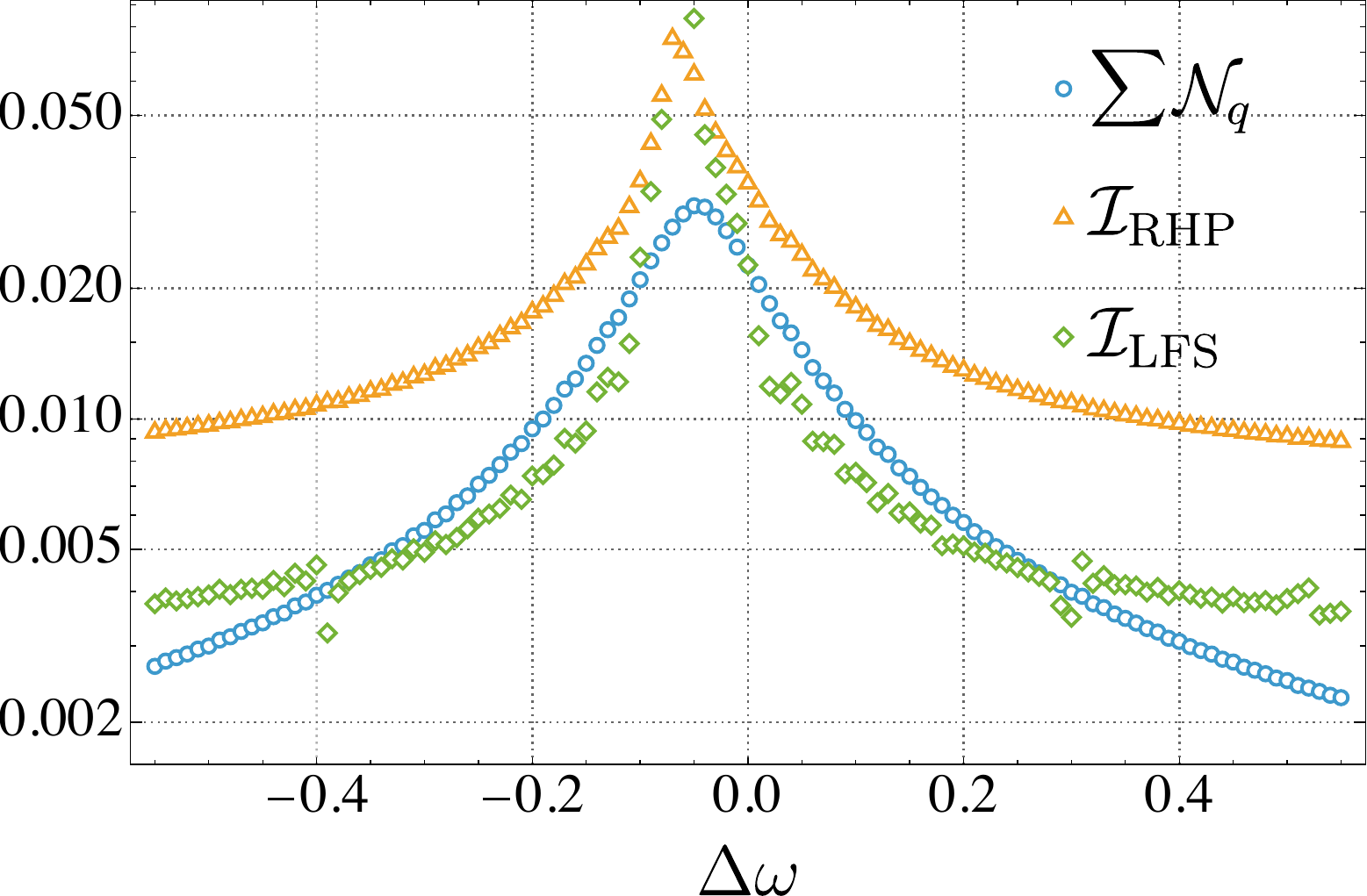}
    \caption{
    Non energy-preserving interactions. 
    {\it Top panel}: Comparison between violations of CP-divisibility identified by $g_n$ (black circles), and $\mathcal N_{\mathrm q}[P_n]$ (brown triangles). We report both the conditions \eqref{eq:CP_conditions} (red dots, green squares and blue rhombs respectively), and the average system's energy-change $\langle\Delta E_S\rangle$ normalized by $\omega_S$ (teal triangles). The system's frequency is $\omega_S = 0.8$ ($\Delta\omega = -0.2$). {\it Bottom panel}: Comparison of $\mathcal{I}_{\text{LFS}}$ (green rhombs), $\mathcal{I}_{\text{RHP}}$ (orange triangles) and $\sum_{n}\mathcal N_{\mathrm q}[P_n]$ (light blue circles), as a function of detuning $\Delta\omega \in [-0.5, 0.5]$ (keeping memory and environment particles at the same frequency). Here, the maximum number of collisions is fixed to $10^3$. All the other parameters are the same as in Fig.~\ref{fig:nonMK-nonpos}.
    }
    \label{fig:NEP}
\end{figure}

{\bf Robustness analysis with non-energy preservation---}When the local energy is not conserved during the dynamics, i.e., $[H_k + H_j, U_{kj}] \neq 0$ with $k,j = S,M,A$, the condition $|c_n|^2 \le a_n(1-b_n)$ is easier to violate; this can occur when considering Heisenberg exchange interactions in the off-resonance regime. This fact is shown in Fig.~\ref{fig:NEP} top panel, where $\omega_S \neq \omega_M = \omega_A$. We denote the system-memory detuning as $\Delta\omega \equiv \omega_S - \omega_M$. In particular, the discrepancy between $g \neq 0$ and $\mathcal N_{\mathrm q}=0$ in the collisions' range $[63, 78]$ is due to the violation of $|c_n|^2 \le a_n(1-b_n)$ while both $a_n, b_n$ belong to $[0,1]$. In the bottom panel of Fig.~\ref{fig:NEP}, we illustrate the behavior of the non-Markovianity measures $\mathcal{I}_{\text{RHP}}, \mathcal{I}_{\text{LFS}}$ as a function of $\Delta\omega$, in comparison with the cumulative non-positivity functional $\sum_{n}\mathcal N_{\mathrm q}[P_n]$. A direct link between memory effects tracked by $\mathcal{I}_{\text{RHP}}, \mathcal{I}_{\text{LFS}}$ and quantum traits of the KDQ statistics of system's energy changes can be observed for all plotted values of $\Delta\omega$. The connection between the two effects is highlighted also by the clear alignment of the peaks of the three plotted quantities; the peaks are all around $\Delta\omega \approx -0.02$. Notably, when the system and the memory are at resonance, the system's energy-change statistics have to be described by a quasiprobability distribution, and this kind of nonclassicality witnesses non-Markovianity.

Some results in the case of anisotropic exchange terms within the interaction Hamiltonian, still leading to $[H_k + H_j, U_{kj}] \neq 0$ ($k,j = S,M,A$), are reported in the Appendix.

{\bf Conclusions---}We determined analytical conditions under which the non-positivity of the KDQ energy-change distribution can witness the non-Markovianity of the system's dynamics. Using the framework of phase covariant maps described by memory-mediated collision models, we showed that the capability for non-Markovianity detection is maximized by dealing with energy-preserving Heisenberg-type exchange interactions. Moreover, whenever the non-positivity of the KDQ distribution witnesses non-Markovianity, anomalous energy fluxes manifest and the QMI becomes non-monotonic. 

As an outlook, one could investigate if a necessary and sufficient condition of non-Markovianity can be given in terms of extended quasiprobabilities, as the ones built with quantum non-demolition measurements~\cite{solinas2024quantum,melegari2025,SolinasReview2026}. It is also certainly interesting to extend our results to other non-Markovian models, leading to decoherence or thermalization, based on a microscopic description of the environment and Hamiltonian interaction terms, as e.g., spin-boson models~\cite{WangPRA2024,SunNatComm2025}, spin qubits~\cite{HaasePRL2018}, integrated quantum dots~\cite{Dalacu_2019,Sapienza2015} or circuit QED systems~\cite{Mirhosseini2019}.   

\begin{acknowledgments}
\textit{Acknowledgments.}---M.P., S.L.~and S.G.~acknowledge financial support from the PRIN project 2022FEXLYB Quantum Reservoir Computing (QuReCo), and the PNRR MUR project PE0000023-NQSTI funded by the European Union---Next Generation EU. 
G.D.C.~acknowledges financial support from the Spanish Agencia
Estatal de Investigación project a PID2024-162153NB-
I00 and UKRI-EPSRC project UKRI3314.
M.P. thanks the support from FCT – Fundação para a Ciência e a Tecnologia (Portugal), namely through project UID/PRR/04540/2025 and contract LA/P/0095/2020.
M.P., G.D.C.~and S.G.~also thank the Royal Society Project IES\textbackslash R3\textbackslash 223086 ``Dissipation-based quantum inference for out-of-equilibrium quantum many-body systems''. A. C. acknowledges support from the UK-EPSRC project EP/W52444X/1.
\end{acknowledgments}

\bibliography{main}

\begin{thebibliography}{52}%
\makeatletter
\providecommand \@ifxundefined [1]{%
 \@ifx{#1\undefined}
}%
\providecommand \@ifnum [1]{%
 \ifnum #1\expandafter \@firstoftwo
 \else \expandafter \@secondoftwo
 \fi
}%
\providecommand \@ifx [1]{%
 \ifx #1\expandafter \@firstoftwo
 \else \expandafter \@secondoftwo
 \fi
}%
\providecommand \natexlab [1]{#1}%
\providecommand \enquote  [1]{``#1''}%
\providecommand \bibnamefont  [1]{#1}%
\providecommand \bibfnamefont [1]{#1}%
\providecommand \citenamefont [1]{#1}%
\providecommand \href@noop [0]{\@secondoftwo}%
\providecommand \href [0]{\begingroup \@sanitize@url \@href}%
\providecommand \@href[1]{\@@startlink{#1}\@@href}%
\providecommand \@@href[1]{\endgroup#1\@@endlink}%
\providecommand \@sanitize@url [0]{\catcode `\\12\catcode `\$12\catcode
  `\&12\catcode `\#12\catcode `\^12\catcode `\_12\catcode `\%12\relax}%
\providecommand \@@startlink[1]{}%
\providecommand \@@endlink[0]{}%
\providecommand \url  [0]{\begingroup\@sanitize@url \@url }%
\providecommand \@url [1]{\endgroup\@href {#1}{\urlprefix }}%
\providecommand \urlprefix  [0]{URL }%
\providecommand \Eprint [0]{\href }%
\providecommand \doibase [0]{https://doi.org/}%
\providecommand \selectlanguage [0]{\@gobble}%
\providecommand \bibinfo  [0]{\@secondoftwo}%
\providecommand \bibfield  [0]{\@secondoftwo}%
\providecommand \translation [1]{[#1]}%
\providecommand \BibitemOpen [0]{}%
\providecommand \bibitemStop [0]{}%
\providecommand \bibitemNoStop [0]{.\EOS\space}%
\providecommand \EOS [0]{\spacefactor3000\relax}%
\providecommand \BibitemShut  [1]{\csname bibitem#1\endcsname}%
\let\auto@bib@innerbib\@empty
\bibitem [{\citenamefont {Altman}\ \emph {et~al.}(2021)\citenamefont {Altman},
  \citenamefont {Brown}, \citenamefont {Carleo}, \citenamefont {Carr},
  \citenamefont {Demler}, \citenamefont {Chin}, \citenamefont {DeMarco},
  \citenamefont {Economou}, \citenamefont {Eriksson}, \citenamefont {Fu},
  \citenamefont {Greiner}, \citenamefont {Hazzard}, \citenamefont {Hulet},
  \citenamefont {Koll\'ar}, \citenamefont {Lev}, \citenamefont {Lukin},
  \citenamefont {Ma}, \citenamefont {Mi}, \citenamefont {Misra}, \citenamefont
  {Monroe}, \citenamefont {Murch}, \citenamefont {Nazario}, \citenamefont {Ni},
  \citenamefont {Potter}, \citenamefont {Roushan}, \citenamefont {Saffman},
  \citenamefont {Schleier-Smith}, \citenamefont {Siddiqi}, \citenamefont
  {Simmonds}, \citenamefont {Singh}, \citenamefont {Spielman}, \citenamefont
  {Temme}, \citenamefont {Weiss}, \citenamefont {Vu\ifmmode \check{c}\else
  \v{c}\fi{}kovi\ifmmode~\acute{c}\else \'{c}\fi{}}, \citenamefont
  {Vuleti\ifmmode~\acute{c}\else \'{c}\fi{}}, \citenamefont {Ye},\ and\
  \citenamefont {Zwierlein}}]{AltmanPRXQ2021}%
  \BibitemOpen
  \bibfield  {author} {\bibinfo {author} {\bibfnamefont {E.}~\bibnamefont
  {Altman}}, \bibinfo {author} {\bibfnamefont {K.~R.}\ \bibnamefont {Brown}},
  \bibinfo {author} {\bibfnamefont {G.}~\bibnamefont {Carleo}}, \bibinfo
  {author} {\bibfnamefont {L.~D.}\ \bibnamefont {Carr}}, \bibinfo {author}
  {\bibfnamefont {E.}~\bibnamefont {Demler}}, \bibinfo {author} {\bibfnamefont
  {C.}~\bibnamefont {Chin}}, \bibinfo {author} {\bibfnamefont {B.}~\bibnamefont
  {DeMarco}}, \bibinfo {author} {\bibfnamefont {S.~E.}\ \bibnamefont
  {Economou}}, \bibinfo {author} {\bibfnamefont {M.~A.}\ \bibnamefont
  {Eriksson}}, \bibinfo {author} {\bibfnamefont {K.-M.~C.}\ \bibnamefont {Fu}},
  \bibinfo {author} {\bibfnamefont {M.}~\bibnamefont {Greiner}}, \bibinfo
  {author} {\bibfnamefont {K.~R.}\ \bibnamefont {Hazzard}}, \bibinfo {author}
  {\bibfnamefont {R.~G.}\ \bibnamefont {Hulet}}, \bibinfo {author}
  {\bibfnamefont {A.~J.}\ \bibnamefont {Koll\'ar}}, \bibinfo {author}
  {\bibfnamefont {B.~L.}\ \bibnamefont {Lev}}, \bibinfo {author} {\bibfnamefont
  {M.~D.}\ \bibnamefont {Lukin}}, \bibinfo {author} {\bibfnamefont
  {R.}~\bibnamefont {Ma}}, \bibinfo {author} {\bibfnamefont {X.}~\bibnamefont
  {Mi}}, \bibinfo {author} {\bibfnamefont {S.}~\bibnamefont {Misra}}, \bibinfo
  {author} {\bibfnamefont {C.}~\bibnamefont {Monroe}}, \bibinfo {author}
  {\bibfnamefont {K.}~\bibnamefont {Murch}}, \bibinfo {author} {\bibfnamefont
  {Z.}~\bibnamefont {Nazario}}, \bibinfo {author} {\bibfnamefont {K.-K.}\
  \bibnamefont {Ni}}, \bibinfo {author} {\bibfnamefont {A.~C.}\ \bibnamefont
  {Potter}}, \bibinfo {author} {\bibfnamefont {P.}~\bibnamefont {Roushan}},
  \bibinfo {author} {\bibfnamefont {M.}~\bibnamefont {Saffman}}, \bibinfo
  {author} {\bibfnamefont {M.}~\bibnamefont {Schleier-Smith}}, \bibinfo
  {author} {\bibfnamefont {I.}~\bibnamefont {Siddiqi}}, \bibinfo {author}
  {\bibfnamefont {R.}~\bibnamefont {Simmonds}}, \bibinfo {author}
  {\bibfnamefont {M.}~\bibnamefont {Singh}}, \bibinfo {author} {\bibfnamefont
  {I.}~\bibnamefont {Spielman}}, \bibinfo {author} {\bibfnamefont
  {K.}~\bibnamefont {Temme}}, \bibinfo {author} {\bibfnamefont {D.~S.}\
  \bibnamefont {Weiss}}, \bibinfo {author} {\bibfnamefont {J.}~\bibnamefont
  {Vu\ifmmode \check{c}\else \v{c}\fi{}kovi\ifmmode~\acute{c}\else
  \'{c}\fi{}}}, \bibinfo {author} {\bibfnamefont {V.}~\bibnamefont
  {Vuleti\ifmmode~\acute{c}\else \'{c}\fi{}}}, \bibinfo {author} {\bibfnamefont
  {J.}~\bibnamefont {Ye}},\ and\ \bibinfo {author} {\bibfnamefont
  {M.}~\bibnamefont {Zwierlein}},\ }\bibfield  {title} {\bibinfo {title}
  {{Quantum Simulators: Architectures and Opportunities}},\ }\href
  {https://doi.org/10.1103/PRXQuantum.2.017003} {\bibfield  {journal} {\bibinfo
   {journal} {PRX Quantum}\ }\textbf {\bibinfo {volume} {2}},\ \bibinfo {pages}
  {017003} (\bibinfo {year} {2021})}\BibitemShut {NoStop}%
\bibitem [{\citenamefont {Rajak}\ \emph {et~al.}(2023)\citenamefont {Rajak},
  \citenamefont {Suzuki}, \citenamefont {Dutta},\ and\ \citenamefont
  {Chakrabarti}}]{Rajak2023}%
  \BibitemOpen
  \bibfield  {author} {\bibinfo {author} {\bibfnamefont {A.}~\bibnamefont
  {Rajak}}, \bibinfo {author} {\bibfnamefont {S.}~\bibnamefont {Suzuki}},
  \bibinfo {author} {\bibfnamefont {A.}~\bibnamefont {Dutta}},\ and\ \bibinfo
  {author} {\bibfnamefont {B.~K.}\ \bibnamefont {Chakrabarti}},\ }\bibfield
  {title} {\bibinfo {title} {Quantum annealing: an overview},\ }\href
  {https://doi.org/10.1098/rsta.2021.0417} {\bibfield  {journal} {\bibinfo
  {journal} {Philos. Trans. A Math. Phys. Eng. Sci.}\ }\textbf {\bibinfo
  {volume} {381}},\ \bibinfo {pages} {20210417} (\bibinfo {year}
  {2023})}\BibitemShut {NoStop}%
\bibitem [{\citenamefont {Ciccarello}\ \emph {et~al.}(2022)\citenamefont
  {Ciccarello}, \citenamefont {Lorenzo}, \citenamefont {Giovannetti},\ and\
  \citenamefont {Palma}}]{ciccarello2022a}%
  \BibitemOpen
  \bibfield  {author} {\bibinfo {author} {\bibfnamefont {F.}~\bibnamefont
  {Ciccarello}}, \bibinfo {author} {\bibfnamefont {S.}~\bibnamefont {Lorenzo}},
  \bibinfo {author} {\bibfnamefont {V.}~\bibnamefont {Giovannetti}},\ and\
  \bibinfo {author} {\bibfnamefont {G.~M.}\ \bibnamefont {Palma}},\ }\bibfield
  {title} {\bibinfo {title} {Quantum collision models: {{Open}} system dynamics
  from repeated interactions},\ }\href
  {https://doi.org/10.1016/j.physrep.2022.01.001} {\bibfield  {journal}
  {\bibinfo  {journal} {Physics Reports}\ }\bibinfo {series} {Quantum Collision
  Models: {{Open}} System Dynamics from Repeated Interactions},\ \textbf
  {\bibinfo {volume} {954}},\ \bibinfo {pages} {1} (\bibinfo {year}
  {2022})}\BibitemShut {NoStop}%
\bibitem [{\citenamefont {Cusumano}(2022)}]{cusimano2022}%
  \BibitemOpen
  \bibfield  {author} {\bibinfo {author} {\bibfnamefont {S.}~\bibnamefont
  {Cusumano}},\ }\bibfield  {title} {\bibinfo {title} {Quantum collision
  models: A beginner guide},\ }\bibfield  {journal} {\bibinfo  {journal}
  {Entropy}\ }\textbf {\bibinfo {volume} {24}},\ \href
  {https://doi.org/10.3390/e24091258} {10.3390/e24091258} (\bibinfo {year}
  {2022})\BibitemShut {NoStop}%
\bibitem [{\citenamefont {Lorenzo}\ \emph {et~al.}(2016)\citenamefont
  {Lorenzo}, \citenamefont {Ciccarello},\ and\ \citenamefont
  {Palma}}]{LorenzoPRA2016}%
  \BibitemOpen
  \bibfield  {author} {\bibinfo {author} {\bibfnamefont {S.}~\bibnamefont
  {Lorenzo}}, \bibinfo {author} {\bibfnamefont {F.}~\bibnamefont
  {Ciccarello}},\ and\ \bibinfo {author} {\bibfnamefont {G.~M.}\ \bibnamefont
  {Palma}},\ }\bibfield  {title} {\bibinfo {title} {Class of exact
  memory-kernel master equations},\ }\href
  {https://doi.org/10.1103/PhysRevA.93.052111} {\bibfield  {journal} {\bibinfo
  {journal} {Phys. Rev. A}\ }\textbf {\bibinfo {volume} {93}},\ \bibinfo
  {pages} {052111} (\bibinfo {year} {2016})}\BibitemShut {NoStop}%
\bibitem [{\citenamefont {Lorenzo}\ \emph {et~al.}(2015)\citenamefont
  {Lorenzo}, \citenamefont {McCloskey}, \citenamefont {Ciccarello},
  \citenamefont {Paternostro},\ and\ \citenamefont {Palma}}]{lorenzo2015b}%
  \BibitemOpen
  \bibfield  {author} {\bibinfo {author} {\bibfnamefont {S.}~\bibnamefont
  {Lorenzo}}, \bibinfo {author} {\bibfnamefont {R.}~\bibnamefont {McCloskey}},
  \bibinfo {author} {\bibfnamefont {F.}~\bibnamefont {Ciccarello}}, \bibinfo
  {author} {\bibfnamefont {M.}~\bibnamefont {Paternostro}},\ and\ \bibinfo
  {author} {\bibfnamefont {G.~M.}\ \bibnamefont {Palma}},\ }\bibfield  {title}
  {\bibinfo {title} {{Landauer's Principle in Multipartite Open Quantum System
  Dynamics}},\ }\href {https://doi.org/10.1103/PhysRevLett.115.120403}
  {\bibfield  {journal} {\bibinfo  {journal} {Phys. Rev. Lett.}\ }\textbf
  {\bibinfo {volume} {115}},\ \bibinfo {pages} {120403} (\bibinfo {year}
  {2015})}\BibitemShut {NoStop}%
\bibitem [{\citenamefont {Lorenzo}\ \emph {et~al.}(2017)\citenamefont
  {Lorenzo}, \citenamefont {Ciccarello}, \citenamefont {Palma},\ and\
  \citenamefont {Vacchini}}]{lorenzo2017g}%
  \BibitemOpen
  \bibfield  {author} {\bibinfo {author} {\bibfnamefont {S.}~\bibnamefont
  {Lorenzo}}, \bibinfo {author} {\bibfnamefont {F.}~\bibnamefont {Ciccarello}},
  \bibinfo {author} {\bibfnamefont {G.~M.}\ \bibnamefont {Palma}},\ and\
  \bibinfo {author} {\bibfnamefont {B.}~\bibnamefont {Vacchini}},\ }\bibfield
  {title} {\bibinfo {title} {Quantum {{Non-Markovian Piecewise Dynamics}} from
  {{Collision Models}}},\ }\href {https://doi.org/10.1142/S123016121740011X}
  {\bibfield  {journal} {\bibinfo  {journal} {Open Systems \& Information
  Dynamics}\ }\textbf {\bibinfo {volume} {24}},\ \bibinfo {pages} {1740011}
  (\bibinfo {year} {2017})}\BibitemShut {NoStop}%
\bibitem [{\citenamefont {Heisenberg}(1928)}]{Heisenberg28}%
  \BibitemOpen
  \bibfield  {author} {\bibinfo {author} {\bibfnamefont {W.}~\bibnamefont
  {Heisenberg}},\ }\bibfield  {title} {\bibinfo {title} {Zur theorie des
  ferromagnetismus - on the theory of ferromagnetism},\ }\href
  {https://doi.org/doi:10.1007/BF01328601} {\bibfield  {journal} {\bibinfo
  {journal} {Zeitschrift für Physik}\ }\textbf {\bibinfo {volume} {49}},\
  \bibinfo {pages} {619} (\bibinfo {year} {1928})}\BibitemShut {NoStop}%
\bibitem [{\citenamefont {Baxter}(1982)}]{Baxter1982}%
  \BibitemOpen
  \bibfield  {author} {\bibinfo {author} {\bibfnamefont {R.~J.}\ \bibnamefont
  {Baxter}},\ }\href@noop {} {\emph {\bibinfo {title} {Exactly solved models in
  statistical mechanics}}}\ (\bibinfo  {publisher} {Academic Press},\ \bibinfo
  {year} {1982})\BibitemShut {NoStop}%
\bibitem [{\citenamefont {Raimond}\ \emph {et~al.}(2001)\citenamefont
  {Raimond}, \citenamefont {Brune},\ and\ \citenamefont
  {Haroche}}]{RaimondRMP2001}%
  \BibitemOpen
  \bibfield  {author} {\bibinfo {author} {\bibfnamefont {J.~M.}\ \bibnamefont
  {Raimond}}, \bibinfo {author} {\bibfnamefont {M.}~\bibnamefont {Brune}},\
  and\ \bibinfo {author} {\bibfnamefont {S.}~\bibnamefont {Haroche}},\
  }\bibfield  {title} {\bibinfo {title} {Manipulating quantum entanglement with
  atoms and photons in a cavity},\ }\href
  {https://doi.org/10.1103/RevModPhys.73.565} {\bibfield  {journal} {\bibinfo
  {journal} {Rev. Mod. Phys.}\ }\textbf {\bibinfo {volume} {73}},\ \bibinfo
  {pages} {565} (\bibinfo {year} {2001})}\BibitemShut {NoStop}%
\bibitem [{\citenamefont {Bernardes}\ \emph {et~al.}(2015)\citenamefont
  {Bernardes}, \citenamefont {Cuevas}, \citenamefont {Orieux}, \citenamefont
  {Monken}, \citenamefont {Mataloni}, \citenamefont {Sciarrino},\ and\
  \citenamefont {Santos}}]{Bernardes2015}%
  \BibitemOpen
  \bibfield  {author} {\bibinfo {author} {\bibfnamefont {N.~K.}\ \bibnamefont
  {Bernardes}}, \bibinfo {author} {\bibfnamefont {A.}~\bibnamefont {Cuevas}},
  \bibinfo {author} {\bibfnamefont {A.}~\bibnamefont {Orieux}}, \bibinfo
  {author} {\bibfnamefont {C.~H.}\ \bibnamefont {Monken}}, \bibinfo {author}
  {\bibfnamefont {P.}~\bibnamefont {Mataloni}}, \bibinfo {author}
  {\bibfnamefont {F.}~\bibnamefont {Sciarrino}},\ and\ \bibinfo {author}
  {\bibfnamefont {M.~F.}\ \bibnamefont {Santos}},\ }\bibfield  {title}
  {\bibinfo {title} {{Experimental observation of weak non-Markovianity}},\
  }\href {https://doi.org/10.1038/srep17520} {\bibfield  {journal} {\bibinfo
  {journal} {Scientific Reports}\ }\textbf {\bibinfo {volume} {5}},\ \bibinfo
  {pages} {17520} (\bibinfo {year} {2015})}\BibitemShut {NoStop}%
\bibitem [{\citenamefont {Maffei}\ \emph {et~al.}(2021)\citenamefont {Maffei},
  \citenamefont {Camati},\ and\ \citenamefont {Auff\`eves}}]{MaffeiPRR2021}%
  \BibitemOpen
  \bibfield  {author} {\bibinfo {author} {\bibfnamefont {M.}~\bibnamefont
  {Maffei}}, \bibinfo {author} {\bibfnamefont {P.~A.}\ \bibnamefont {Camati}},\
  and\ \bibinfo {author} {\bibfnamefont {A.}~\bibnamefont {Auff\`eves}},\
  }\bibfield  {title} {\bibinfo {title} {Probing nonclassical light fields with
  energetic witnesses in waveguide quantum electrodynamics},\ }\href
  {https://doi.org/10.1103/PhysRevResearch.3.L032073} {\bibfield  {journal}
  {\bibinfo  {journal} {Phys. Rev. Res.}\ }\textbf {\bibinfo {volume} {3}},\
  \bibinfo {pages} {L032073} (\bibinfo {year} {2021})}\BibitemShut {NoStop}%
\bibitem [{\citenamefont {Filippov}\ \emph {et~al.}(2020)\citenamefont
  {Filippov}, \citenamefont {Glinov},\ and\ \citenamefont
  {Lepp{\"a}j{\"a}rvi}}]{filippov2020}%
  \BibitemOpen
  \bibfield  {author} {\bibinfo {author} {\bibfnamefont {S.~N.}\ \bibnamefont
  {Filippov}}, \bibinfo {author} {\bibfnamefont {A.~N.}\ \bibnamefont
  {Glinov}},\ and\ \bibinfo {author} {\bibfnamefont {L.}~\bibnamefont
  {Lepp{\"a}j{\"a}rvi}},\ }\bibfield  {title} {\bibinfo {title} {Phase
  covariant qubit dynamics and divisibility},\ }\href
  {https://doi.org/10.1134/S1995080220040095} {\bibfield  {journal} {\bibinfo
  {journal} {Lobachevskii Journal of Mathematics}\ }\textbf {\bibinfo {volume}
  {41}},\ \bibinfo {pages} {617} (\bibinfo {year} {2020})}\BibitemShut
  {NoStop}%
\bibitem [{\citenamefont {Smirne}\ \emph {et~al.}(2016)\citenamefont {Smirne},
  \citenamefont {Ko\l{}ody\ifmmode~\acute{n}\else \'{n}\fi{}ski}, \citenamefont
  {Huelga},\ and\ \citenamefont {Demkowicz-Dobrza\ifmmode~\acute{n}\else
  \'{n}\fi{}ski}}]{Smirne2016}%
  \BibitemOpen
  \bibfield  {author} {\bibinfo {author} {\bibfnamefont {A.}~\bibnamefont
  {Smirne}}, \bibinfo {author} {\bibfnamefont {J.}~\bibnamefont
  {Ko\l{}ody\ifmmode~\acute{n}\else \'{n}\fi{}ski}}, \bibinfo {author}
  {\bibfnamefont {S.~F.}\ \bibnamefont {Huelga}},\ and\ \bibinfo {author}
  {\bibfnamefont {R.}~\bibnamefont {Demkowicz-Dobrza\ifmmode~\acute{n}\else
  \'{n}\fi{}ski}},\ }\bibfield  {title} {\bibinfo {title} {Ultimate precision
  limits for noisy frequency estimation},\ }\href
  {https://doi.org/10.1103/PhysRevLett.116.120801} {\bibfield  {journal}
  {\bibinfo  {journal} {Phys. Rev. Lett.}\ }\textbf {\bibinfo {volume} {116}},\
  \bibinfo {pages} {120801} (\bibinfo {year} {2016})}\BibitemShut {NoStop}%
\bibitem [{\citenamefont {Siudzińska}(2023)}]{siudzinska2023}%
  \BibitemOpen
  \bibfield  {author} {\bibinfo {author} {\bibfnamefont {K.}~\bibnamefont
  {Siudzińska}},\ }\bibfield  {title} {\bibinfo {title} {Geometry of
  phase-covariant qubit channels},\ }\href
  {https://doi.org/10.1088/2399-6528/ace0f4} {\bibfield  {journal} {\bibinfo
  {journal} {J. Phys. Commun.}\ }\textbf {\bibinfo {volume} {7}},\ \bibinfo
  {pages} {075002} (\bibinfo {year} {2023})}\BibitemShut {NoStop}%
\bibitem [{\citenamefont {Holevo}(1993)}]{Holevo1993}%
  \BibitemOpen
  \bibfield  {author} {\bibinfo {author} {\bibfnamefont {A.}~\bibnamefont
  {Holevo}},\ }\bibfield  {title} {\bibinfo {title} {A note on covariant
  dynamical semigroups},\ }\href
  {https://doi.org/https://doi.org/10.1016/0034-4877(93)90014-6} {\bibfield
  {journal} {\bibinfo  {journal} {Rep. Math. Phys.}\ }\textbf {\bibinfo
  {volume} {32}},\ \bibinfo {pages} {211} (\bibinfo {year} {1993})}\BibitemShut
  {NoStop}%
\bibitem [{\citenamefont {Rivas}\ \emph {et~al.}(2014)\citenamefont {Rivas},
  \citenamefont {Huelga},\ and\ \citenamefont {Plenio}}]{rivas_2014}%
  \BibitemOpen
  \bibfield  {author} {\bibinfo {author} {\bibfnamefont {{\'A}.}~\bibnamefont
  {Rivas}}, \bibinfo {author} {\bibfnamefont {S.~F.}\ \bibnamefont {Huelga}},\
  and\ \bibinfo {author} {\bibfnamefont {M.~B.}\ \bibnamefont {Plenio}},\
  }\bibfield  {title} {\bibinfo {title} {Quantum non-{{Markovianity}}:
  Characterization, quantification and detection},\ }\href
  {https://doi.org/10.1088/0034-4885/77/9/094001} {\bibfield  {journal}
  {\bibinfo  {journal} {Reports on Progress in Physics}\ }\textbf {\bibinfo
  {volume} {77}},\ \bibinfo {pages} {094001} (\bibinfo {year}
  {2014})}\BibitemShut {NoStop}%
\bibitem [{\citenamefont {Shrikant}\ and\ \citenamefont
  {Mandayam}(2023)}]{shrikant2023}%
  \BibitemOpen
  \bibfield  {author} {\bibinfo {author} {\bibfnamefont {U.}~\bibnamefont
  {Shrikant}}\ and\ \bibinfo {author} {\bibfnamefont {P.}~\bibnamefont
  {Mandayam}},\ }\bibfield  {title} {\bibinfo {title} {Quantum
  non-{{Markovianity}}: {{Overview}} and recent developments},\ }\bibfield
  {journal} {\bibinfo  {journal} {Front. Quantum Sci. Technol.}\ }\textbf
  {\bibinfo {volume} {2}},\ \href {https://doi.org/10.3389/frqst.2023.1134583}
  {10.3389/frqst.2023.1134583} (\bibinfo {year} {2023})\BibitemShut {NoStop}%
\bibitem [{\citenamefont {Ciccarello}(2017)}]{Ciccarello2017}%
  \BibitemOpen
  \bibfield  {author} {\bibinfo {author} {\bibfnamefont {F.}~\bibnamefont
  {Ciccarello}},\ }\bibfield  {title} {\bibinfo {title} {Collision models in
  quantum optics},\ }\bibfield  {journal} {\bibinfo  {journal} {Quantum
  Measurements and Quantum Metrology}\ }\textbf {\bibinfo {volume} {4}},\ \href
  {https://doi.org/10.1515/qmetro-2017-0007} {10.1515/qmetro-2017-0007}
  (\bibinfo {year} {2017})\BibitemShut {NoStop}%
\bibitem [{\citenamefont {Meschede}\ \emph {et~al.}(1985)\citenamefont
  {Meschede}, \citenamefont {Walther},\ and\ \citenamefont
  {M\"uller}}]{Meschede1985}%
  \BibitemOpen
  \bibfield  {author} {\bibinfo {author} {\bibfnamefont {D.}~\bibnamefont
  {Meschede}}, \bibinfo {author} {\bibfnamefont {H.}~\bibnamefont {Walther}},\
  and\ \bibinfo {author} {\bibfnamefont {G.}~\bibnamefont {M\"uller}},\
  }\bibfield  {title} {\bibinfo {title} {One-atom maser},\ }\href
  {https://doi.org/10.1103/PhysRevLett.54.551} {\bibfield  {journal} {\bibinfo
  {journal} {Phys. Rev. Lett.}\ }\textbf {\bibinfo {volume} {54}},\ \bibinfo
  {pages} {551} (\bibinfo {year} {1985})}\BibitemShut {NoStop}%
\bibitem [{\citenamefont {Ferreira}\ \emph {et~al.}(2021)\citenamefont
  {Ferreira}, \citenamefont {Banker}, \citenamefont {Sipahigil}, \citenamefont
  {Matheny}, \citenamefont {Keller}, \citenamefont {Kim}, \citenamefont
  {Mirhosseini},\ and\ \citenamefont {Painter}}]{Ferreira2021}%
  \BibitemOpen
  \bibfield  {author} {\bibinfo {author} {\bibfnamefont {V.~S.}\ \bibnamefont
  {Ferreira}}, \bibinfo {author} {\bibfnamefont {J.}~\bibnamefont {Banker}},
  \bibinfo {author} {\bibfnamefont {A.}~\bibnamefont {Sipahigil}}, \bibinfo
  {author} {\bibfnamefont {M.~H.}\ \bibnamefont {Matheny}}, \bibinfo {author}
  {\bibfnamefont {A.~J.}\ \bibnamefont {Keller}}, \bibinfo {author}
  {\bibfnamefont {E.}~\bibnamefont {Kim}}, \bibinfo {author} {\bibfnamefont
  {M.}~\bibnamefont {Mirhosseini}},\ and\ \bibinfo {author} {\bibfnamefont
  {O.}~\bibnamefont {Painter}},\ }\bibfield  {title} {\bibinfo {title}
  {{Collapse and Revival of an Artificial Atom Coupled to a Structured Photonic
  Reservoir}},\ }\href {https://doi.org/10.1103/PhysRevX.11.041043} {\bibfield
  {journal} {\bibinfo  {journal} {Phys. Rev. X}\ }\textbf {\bibinfo {volume}
  {11}},\ \bibinfo {pages} {041043} (\bibinfo {year} {2021})}\BibitemShut
  {NoStop}%
\bibitem [{\citenamefont {Cattaneo}\ \emph {et~al.}(2021)\citenamefont
  {Cattaneo}, \citenamefont {De~Chiara}, \citenamefont {Maniscalco},
  \citenamefont {Zambrini},\ and\ \citenamefont {Giorgi}}]{CattaneoPRL2021}%
  \BibitemOpen
  \bibfield  {author} {\bibinfo {author} {\bibfnamefont {M.}~\bibnamefont
  {Cattaneo}}, \bibinfo {author} {\bibfnamefont {G.}~\bibnamefont {De~Chiara}},
  \bibinfo {author} {\bibfnamefont {S.}~\bibnamefont {Maniscalco}}, \bibinfo
  {author} {\bibfnamefont {R.}~\bibnamefont {Zambrini}},\ and\ \bibinfo
  {author} {\bibfnamefont {G.~L.}\ \bibnamefont {Giorgi}},\ }\bibfield  {title}
  {\bibinfo {title} {{Collision Models Can Efficiently Simulate Any
  Multipartite Markovian Quantum Dynamics}},\ }\href
  {https://doi.org/10.1103/PhysRevLett.126.130403} {\bibfield  {journal}
  {\bibinfo  {journal} {Phys. Rev. Lett.}\ }\textbf {\bibinfo {volume} {126}},\
  \bibinfo {pages} {130403} (\bibinfo {year} {2021})}\BibitemShut {NoStop}%
\bibitem [{\citenamefont {Lacroix}\ \emph {et~al.}(2025)\citenamefont
  {Lacroix}, \citenamefont {Cilluffo}, \citenamefont {Huelga},\ and\
  \citenamefont {Plenio}}]{Lacroix2025}%
  \BibitemOpen
  \bibfield  {author} {\bibinfo {author} {\bibfnamefont {T.}~\bibnamefont
  {Lacroix}}, \bibinfo {author} {\bibfnamefont {D.}~\bibnamefont {Cilluffo}},
  \bibinfo {author} {\bibfnamefont {S.~F.}\ \bibnamefont {Huelga}},\ and\
  \bibinfo {author} {\bibfnamefont {M.~B.}\ \bibnamefont {Plenio}},\ }\bibfield
   {title} {\bibinfo {title} {Making quantum collision models exact},\ }\href
  {https://doi.org/10.1038/s42005-025-02201-2} {\bibfield  {journal} {\bibinfo
  {journal} {Communications Physics}\ }\textbf {\bibinfo {volume} {8}},\
  \bibinfo {pages} {268} (\bibinfo {year} {2025})}\BibitemShut {NoStop}%
\bibitem [{\citenamefont {Levy}\ and\ \citenamefont
  {Lostaglio}(2020)}]{LevyPRX2020}%
  \BibitemOpen
  \bibfield  {author} {\bibinfo {author} {\bibfnamefont {A.}~\bibnamefont
  {Levy}}\ and\ \bibinfo {author} {\bibfnamefont {M.}~\bibnamefont
  {Lostaglio}},\ }\bibfield  {title} {\bibinfo {title} {{Quasiprobability
  Distribution for Heat Fluctuations in the Quantum Regime}},\ }\href
  {https://doi.org/10.1103/PRXQuantum.1.010309} {\bibfield  {journal} {\bibinfo
   {journal} {PRX Quantum}\ }\textbf {\bibinfo {volume} {1}},\ \bibinfo {pages}
  {010309} (\bibinfo {year} {2020})}\BibitemShut {NoStop}%
\bibitem [{\citenamefont {Lostaglio}\ \emph {et~al.}(2023)\citenamefont
  {Lostaglio}, \citenamefont {Belenchia}, \citenamefont {Levy}, \citenamefont
  {Hernandez-Gomez}, \citenamefont {Fabbri},\ and\ \citenamefont
  {Gherardini}}]{LostaglioQuantum2023}%
  \BibitemOpen
  \bibfield  {author} {\bibinfo {author} {\bibfnamefont {M.}~\bibnamefont
  {Lostaglio}}, \bibinfo {author} {\bibfnamefont {A.}~\bibnamefont
  {Belenchia}}, \bibinfo {author} {\bibfnamefont {A.}~\bibnamefont {Levy}},
  \bibinfo {author} {\bibfnamefont {S.}~\bibnamefont {Hernandez-Gomez}},
  \bibinfo {author} {\bibfnamefont {N.}~\bibnamefont {Fabbri}},\ and\ \bibinfo
  {author} {\bibfnamefont {S.}~\bibnamefont {Gherardini}},\ }\bibfield  {title}
  {\bibinfo {title} {{Kirkwood-Dirac quasiprobability approach to the
  statistics of incompatible observables}},\ }\href
  {https://doi.org/10.22331/q-2023-10-09-1128} {\bibfield  {journal} {\bibinfo
  {journal} {Quantum}\ }\textbf {\bibinfo {volume} {7}},\ \bibinfo {pages}
  {1128} (\bibinfo {year} {2023})}\BibitemShut {NoStop}%
\bibitem [{\citenamefont {Hern\'andez-G\'omez}\ \emph
  {et~al.}(2024)\citenamefont {Hern\'andez-G\'omez}, \citenamefont
  {Gherardini}, \citenamefont {Belenchia}, \citenamefont {Lostaglio},
  \citenamefont {Levy},\ and\ \citenamefont
  {Fabbri}}]{hernandez2022experimental}%
  \BibitemOpen
  \bibfield  {author} {\bibinfo {author} {\bibfnamefont {S.}~\bibnamefont
  {Hern\'andez-G\'omez}}, \bibinfo {author} {\bibfnamefont {S.}~\bibnamefont
  {Gherardini}}, \bibinfo {author} {\bibfnamefont {A.}~\bibnamefont
  {Belenchia}}, \bibinfo {author} {\bibfnamefont {M.}~\bibnamefont
  {Lostaglio}}, \bibinfo {author} {\bibfnamefont {A.}~\bibnamefont {Levy}},\
  and\ \bibinfo {author} {\bibfnamefont {N.}~\bibnamefont {Fabbri}},\
  }\bibfield  {title} {\bibinfo {title} {Projective measurements can probe
  nonclassical work extraction and time correlations},\ }\href
  {https://doi.org/10.1103/PhysRevResearch.6.023280} {\bibfield  {journal}
  {\bibinfo  {journal} {Phys. Rev. Res.}\ }\textbf {\bibinfo {volume} {6}},\
  \bibinfo {pages} {023280} (\bibinfo {year} {2024})}\BibitemShut {NoStop}%
\bibitem [{\citenamefont {Santini}\ \emph {et~al.}(2023)\citenamefont
  {Santini}, \citenamefont {Solfanelli}, \citenamefont {Gherardini},\ and\
  \citenamefont {Collura}}]{SantiniPRB2023}%
  \BibitemOpen
  \bibfield  {author} {\bibinfo {author} {\bibfnamefont {A.}~\bibnamefont
  {Santini}}, \bibinfo {author} {\bibfnamefont {A.}~\bibnamefont {Solfanelli}},
  \bibinfo {author} {\bibfnamefont {S.}~\bibnamefont {Gherardini}},\ and\
  \bibinfo {author} {\bibfnamefont {M.}~\bibnamefont {Collura}},\ }\bibfield
  {title} {\bibinfo {title} {Work statistics, quantum signatures, and enhanced
  work extraction in quadratic fermionic models},\ }\href
  {https://doi.org/10.1103/PhysRevB.108.104308} {\bibfield  {journal} {\bibinfo
   {journal} {Phys. Rev. B}\ }\textbf {\bibinfo {volume} {108}},\ \bibinfo
  {pages} {104308} (\bibinfo {year} {2023})}\BibitemShut {NoStop}%
\bibitem [{\citenamefont {Gherardini}\ and\ \citenamefont
  {De~Chiara}(2024)}]{GherardiniTutorial}%
  \BibitemOpen
  \bibfield  {author} {\bibinfo {author} {\bibfnamefont {S.}~\bibnamefont
  {Gherardini}}\ and\ \bibinfo {author} {\bibfnamefont {G.}~\bibnamefont
  {De~Chiara}},\ }\bibfield  {title} {\bibinfo {title} {{Quasiprobabilities in
  Quantum Thermodynamics and Many-Body Systems}},\ }\href
  {https://doi.org/10.1103/PRXQuantum.5.030201} {\bibfield  {journal} {\bibinfo
   {journal} {PRX Quantum}\ }\textbf {\bibinfo {volume} {5}},\ \bibinfo {pages}
  {030201} (\bibinfo {year} {2024})}\BibitemShut {NoStop}%
\bibitem [{\citenamefont {Hern{\'a}ndez-G{\'o}mez}\ \emph
  {et~al.}(2024)\citenamefont {Hern{\'a}ndez-G{\'o}mez}, \citenamefont
  {Isogawa}, \citenamefont {Belenchia}, \citenamefont {Levy}, \citenamefont
  {Fabbri}, \citenamefont {Gherardini},\ and\ \citenamefont
  {Cappellaro}}]{HernandezNpjQI2024}%
  \BibitemOpen
  \bibfield  {author} {\bibinfo {author} {\bibfnamefont {S.}~\bibnamefont
  {Hern{\'a}ndez-G{\'o}mez}}, \bibinfo {author} {\bibfnamefont
  {T.}~\bibnamefont {Isogawa}}, \bibinfo {author} {\bibfnamefont
  {A.}~\bibnamefont {Belenchia}}, \bibinfo {author} {\bibfnamefont
  {A.}~\bibnamefont {Levy}}, \bibinfo {author} {\bibfnamefont {N.}~\bibnamefont
  {Fabbri}}, \bibinfo {author} {\bibfnamefont {S.}~\bibnamefont {Gherardini}},\
  and\ \bibinfo {author} {\bibfnamefont {P.}~\bibnamefont {Cappellaro}},\
  }\bibfield  {title} {\bibinfo {title} {Interferometry of quantum correlation
  functions to access quasiprobability distribution of work},\ }\bibfield
  {journal} {\bibinfo  {journal} {npj Quantum Information}\ }\textbf {\bibinfo
  {volume} {10}},\ \href {https://doi.org/10.1038/s41534-024-00913-x}
  {10.1038/s41534-024-00913-x} (\bibinfo {year} {2024})\BibitemShut {NoStop}%
\bibitem [{\citenamefont {Arvidsson-Shukur}\ \emph {et~al.}(2024)\citenamefont
  {Arvidsson-Shukur}, \citenamefont {Braasch~Jr.}, \citenamefont {De~Bievre},
  \citenamefont {Dressel}, \citenamefont {Jordan}, \citenamefont {Langrenez},
  \citenamefont {Lostaglio}, \citenamefont {Lundeen},\ and\ \citenamefont
  {Yunger~Halpern}}]{ArvidssonShukur2024review}%
  \BibitemOpen
  \bibfield  {author} {\bibinfo {author} {\bibfnamefont {D.~R.~M.}\
  \bibnamefont {Arvidsson-Shukur}}, \bibinfo {author} {\bibfnamefont {W.~F.}\
  \bibnamefont {Braasch~Jr.}}, \bibinfo {author} {\bibfnamefont
  {S.}~\bibnamefont {De~Bievre}}, \bibinfo {author} {\bibfnamefont
  {J.}~\bibnamefont {Dressel}}, \bibinfo {author} {\bibfnamefont {A.~N.}\
  \bibnamefont {Jordan}}, \bibinfo {author} {\bibfnamefont {C.}~\bibnamefont
  {Langrenez}}, \bibinfo {author} {\bibfnamefont {M.}~\bibnamefont
  {Lostaglio}}, \bibinfo {author} {\bibfnamefont {J.~S.}\ \bibnamefont
  {Lundeen}},\ and\ \bibinfo {author} {\bibfnamefont {N.}~\bibnamefont
  {Yunger~Halpern}},\ }\bibfield  {title} {\bibinfo {title} {Properties and
  applications of the kirkwood-dirac distribution},\ }\href
  {https://doi.org/10.1088/1367-2630/ada05d} {\bibfield  {journal} {\bibinfo
  {journal} {New J. Phys.}\ }\textbf {\bibinfo {volume} {26}},\ \bibinfo
  {pages} {121201} (\bibinfo {year} {2024})}\BibitemShut {NoStop}%
\bibitem [{\citenamefont {Pezzutto}\ \emph {et~al.}(2025)\citenamefont
  {Pezzutto}, \citenamefont {De~Chiara},\ and\ \citenamefont
  {Gherardini}}]{Pezzutto2025non-positive}%
  \BibitemOpen
  \bibfield  {author} {\bibinfo {author} {\bibfnamefont {M.}~\bibnamefont
  {Pezzutto}}, \bibinfo {author} {\bibfnamefont {G.}~\bibnamefont
  {De~Chiara}},\ and\ \bibinfo {author} {\bibfnamefont {S.}~\bibnamefont
  {Gherardini}},\ }\bibfield  {title} {\bibinfo {title} {Non-positive energy
  quasidistributions in coherent collision models},\ }\href
  {https://doi.org/10.1088/2058-9565/aded2e} {\bibfield  {journal} {\bibinfo
  {journal} {Quantum Sci. Technol.}\ }\textbf {\bibinfo {volume} {10}},\
  \bibinfo {pages} {035066} (\bibinfo {year} {2025})}\BibitemShut {NoStop}%
\bibitem [{\citenamefont {Yoshimura}\ and\ \citenamefont
  {Hamazaki}(2026)}]{Yoshimura2025}%
  \BibitemOpen
  \bibfield  {author} {\bibinfo {author} {\bibfnamefont {K.}~\bibnamefont
  {Yoshimura}}\ and\ \bibinfo {author} {\bibfnamefont {R.}~\bibnamefont
  {Hamazaki}},\ }\bibfield  {title} {\bibinfo {title} {Quasiprobability
  thermodynamic uncertainty relation},\ }\href
  {https://doi.org/10.1103/ky8n-9bcy} {\bibfield  {journal} {\bibinfo
  {journal} {Phys. Rev. Lett.}\ }\textbf {\bibinfo {volume} {136}},\ \bibinfo
  {pages} {120406} (\bibinfo {year} {2026})}\BibitemShut {NoStop}%
\bibitem [{\citenamefont {Perciavalle}\ \emph {et~al.}(2026)\citenamefont
  {Perciavalle}, \citenamefont {Lo~Gullo},\ and\ \citenamefont
  {Plastina}}]{Perciavalle2025}%
  \BibitemOpen
  \bibfield  {author} {\bibinfo {author} {\bibfnamefont {F.}~\bibnamefont
  {Perciavalle}}, \bibinfo {author} {\bibfnamefont {N.}~\bibnamefont
  {Lo~Gullo}},\ and\ \bibinfo {author} {\bibfnamefont {F.}~\bibnamefont
  {Plastina}},\ }\bibfield  {title} {\bibinfo {title} {Quantum coherence and
  anomalous work extraction in qubit gate dynamics},\ }\href
  {https://doi.org/10.1103/qxmt-m3rk} {\bibfield  {journal} {\bibinfo
  {journal} {Phys. Rev. E}\ }\textbf {\bibinfo {volume} {113}},\ \bibinfo
  {pages} {054126} (\bibinfo {year} {2026})}\BibitemShut {NoStop}%
\bibitem [{\citenamefont {Donelli}\ \emph {et~al.}(2026)\citenamefont
  {Donelli}, \citenamefont {De~Chiara}, \citenamefont {Scazza},\ and\
  \citenamefont {Gherardini}}]{DonelliPRA2026}%
  \BibitemOpen
  \bibfield  {author} {\bibinfo {author} {\bibfnamefont {B.}~\bibnamefont
  {Donelli}}, \bibinfo {author} {\bibfnamefont {G.}~\bibnamefont {De~Chiara}},
  \bibinfo {author} {\bibfnamefont {F.}~\bibnamefont {Scazza}},\ and\ \bibinfo
  {author} {\bibfnamefont {S.}~\bibnamefont {Gherardini}},\ }\bibfield  {title}
  {\bibinfo {title} {Impact of quantum coherence on the dynamics and
  thermodynamics of quenched free fermions coupled to a localized defect},\
  }\href {https://doi.org/10.1103/5kt1-5m83} {\bibfield  {journal} {\bibinfo
  {journal} {Phys. Rev. A}\ }\textbf {\bibinfo {volume} {113}},\ \bibinfo
  {pages} {013311} (\bibinfo {year} {2026})}\BibitemShut {NoStop}%
\bibitem [{\citenamefont {Rivas}\ \emph {et~al.}(2010)\citenamefont {Rivas},
  \citenamefont {Huelga},\ and\ \citenamefont {Plenio}}]{RHP-non-Markovianity}%
  \BibitemOpen
  \bibfield  {author} {\bibinfo {author} {\bibfnamefont {A.}~\bibnamefont
  {Rivas}}, \bibinfo {author} {\bibfnamefont {S.~F.}\ \bibnamefont {Huelga}},\
  and\ \bibinfo {author} {\bibfnamefont {M.~B.}\ \bibnamefont {Plenio}},\
  }\bibfield  {title} {\bibinfo {title} {{Entanglement and Non-Markovianity of
  Quantum Evolutions}},\ }\href
  {https://doi.org/10.1103/PhysRevLett.105.050403} {\bibfield  {journal}
  {\bibinfo  {journal} {Phys. Rev. Lett.}\ }\textbf {\bibinfo {volume} {105}},\
  \bibinfo {pages} {050403} (\bibinfo {year} {2010})}\BibitemShut {NoStop}%
\bibitem [{\citenamefont {Luo}\ \emph {et~al.}(2012{\natexlab{a}})\citenamefont
  {Luo}, \citenamefont {Fu},\ and\ \citenamefont {Song}}]{LuoPRA2012}%
  \BibitemOpen
  \bibfield  {author} {\bibinfo {author} {\bibfnamefont {S.}~\bibnamefont
  {Luo}}, \bibinfo {author} {\bibfnamefont {S.}~\bibnamefont {Fu}},\ and\
  \bibinfo {author} {\bibfnamefont {H.}~\bibnamefont {Song}},\ }\bibfield
  {title} {\bibinfo {title} {{Quantifying non-Markovianity via correlations}},\
  }\href {https://doi.org/10.1103/PhysRevA.86.044101} {\bibfield  {journal}
  {\bibinfo  {journal} {Phys. Rev. A}\ }\textbf {\bibinfo {volume} {86}},\
  \bibinfo {pages} {044101} (\bibinfo {year} {2012}{\natexlab{a}})}\BibitemShut
  {NoStop}%
\bibitem [{\citenamefont {Solinas}\ \emph {et~al.}(2022)\citenamefont
  {Solinas}, \citenamefont {Amico},\ and\ \citenamefont
  {Zangh\`{\i}}}]{solinas2022}%
  \BibitemOpen
  \bibfield  {author} {\bibinfo {author} {\bibfnamefont {P.}~\bibnamefont
  {Solinas}}, \bibinfo {author} {\bibfnamefont {M.}~\bibnamefont {Amico}},\
  and\ \bibinfo {author} {\bibfnamefont {N.}~\bibnamefont {Zangh\`{\i}}},\
  }\bibfield  {title} {\bibinfo {title} {Quasiprobabilities of work and heat in
  an open quantum system},\ }\href
  {https://doi.org/10.1103/PhysRevA.105.032606} {\bibfield  {journal} {\bibinfo
   {journal} {Phys. Rev. A}\ }\textbf {\bibinfo {volume} {105}},\ \bibinfo
  {pages} {032606} (\bibinfo {year} {2022})}\BibitemShut {NoStop}%
\bibitem [{\citenamefont {Comar}\ \emph {et~al.}(2025)\citenamefont {Comar},
  \citenamefont {Cius}, \citenamefont {Santos}, \citenamefont {Wagner},\ and\
  \citenamefont {Amaral}}]{ComarPRXQuantum2025}%
  \BibitemOpen
  \bibfield  {author} {\bibinfo {author} {\bibfnamefont {N.~E.}\ \bibnamefont
  {Comar}}, \bibinfo {author} {\bibfnamefont {D.}~\bibnamefont {Cius}},
  \bibinfo {author} {\bibfnamefont {L.~F.}\ \bibnamefont {Santos}}, \bibinfo
  {author} {\bibfnamefont {R.}~\bibnamefont {Wagner}},\ and\ \bibinfo {author}
  {\bibfnamefont {B.}~\bibnamefont {Amaral}},\ }\bibfield  {title} {\bibinfo
  {title} {{Contextuality in Anomalous Heat Flow}},\ }\href
  {https://doi.org/10.1103/f68k-cjx4} {\bibfield  {journal} {\bibinfo
  {journal} {PRX Quantum}\ }\textbf {\bibinfo {volume} {6}},\ \bibinfo {pages}
  {030359} (\bibinfo {year} {2025})}\BibitemShut {NoStop}%
\bibitem [{\citenamefont {Mallik}\ \emph {et~al.}(2025)\citenamefont {Mallik},
  \citenamefont {Cangemi}, \citenamefont {Levy},\ and\ \citenamefont
  {Dalla~Torre}}]{Mallik2025}%
  \BibitemOpen
  \bibfield  {author} {\bibinfo {author} {\bibfnamefont {A.~V.}\ \bibnamefont
  {Mallik}}, \bibinfo {author} {\bibfnamefont {L.~M.}\ \bibnamefont {Cangemi}},
  \bibinfo {author} {\bibfnamefont {A.}~\bibnamefont {Levy}},\ and\ \bibinfo
  {author} {\bibfnamefont {E.~G.}\ \bibnamefont {Dalla~Torre}},\ }\bibfield
  {title} {\bibinfo {title} {Probing quantum anomalous heat flow using
  mid-circuit measurements},\ }\href {https://doi.org/10.1002/qute.202500328}
  {\bibfield  {journal} {\bibinfo  {journal} {Adv. Quantum Technol.}\ }\textbf
  {\bibinfo {volume} {8}},\ \bibinfo {pages} {e00328} (\bibinfo {year}
  {2025})}\BibitemShut {NoStop}%
\bibitem [{\citenamefont {Huang}\ \emph {et~al.}(2026)\citenamefont {Huang},
  \citenamefont {Zhang}, \citenamefont {Liu}, \citenamefont {Li}, \citenamefont
  {Long}, \citenamefont {Liu}, \citenamefont {Wang}, \citenamefont {Fan},
  \citenamefont {Zheng}, \citenamefont {Feng}, \citenamefont {Zhou},
  \citenamefont {Ng}, \citenamefont {Nie}, \citenamefont {Man},\ and\
  \citenamefont {Lu}}]{HuangPRL2026}%
  \BibitemOpen
  \bibfield  {author} {\bibinfo {author} {\bibfnamefont {K.}~\bibnamefont
  {Huang}}, \bibinfo {author} {\bibfnamefont {Q.}~\bibnamefont {Zhang}},
  \bibinfo {author} {\bibfnamefont {X.}~\bibnamefont {Liu}}, \bibinfo {author}
  {\bibfnamefont {R.}~\bibnamefont {Li}}, \bibinfo {author} {\bibfnamefont
  {X.}~\bibnamefont {Long}}, \bibinfo {author} {\bibfnamefont {H.}~\bibnamefont
  {Liu}}, \bibinfo {author} {\bibfnamefont {X.}~\bibnamefont {Wang}}, \bibinfo
  {author} {\bibfnamefont {Y.-a.}\ \bibnamefont {Fan}}, \bibinfo {author}
  {\bibfnamefont {Y.}~\bibnamefont {Zheng}}, \bibinfo {author} {\bibfnamefont
  {Y.}~\bibnamefont {Feng}}, \bibinfo {author} {\bibfnamefont {Y.}~\bibnamefont
  {Zhou}}, \bibinfo {author} {\bibfnamefont {J.}~\bibnamefont {Ng}}, \bibinfo
  {author} {\bibfnamefont {X.}~\bibnamefont {Nie}}, \bibinfo {author}
  {\bibfnamefont {Z.-X.}\ \bibnamefont {Man}},\ and\ \bibinfo {author}
  {\bibfnamefont {D.}~\bibnamefont {Lu}},\ }\bibfield  {title} {\bibinfo
  {title} {{Reversing Heat Flow by Coherence in a Multipartite Quantum
  System}},\ }\href {https://doi.org/10.1103/sjvr-fdvh} {\bibfield  {journal}
  {\bibinfo  {journal} {Phys. Rev. Lett.}\ }\textbf {\bibinfo {volume} {136}},\
  \bibinfo {pages} {050403} (\bibinfo {year} {2026})}\BibitemShut {NoStop}%
\bibitem [{\citenamefont {Li}\ \emph {et~al.}(2024)\citenamefont {Li},
  \citenamefont {Zou},\ and\ \citenamefont {Shao}}]{LiPRA2024}%
  \BibitemOpen
  \bibfield  {author} {\bibinfo {author} {\bibfnamefont {H.}~\bibnamefont
  {Li}}, \bibinfo {author} {\bibfnamefont {J.}~\bibnamefont {Zou}},\ and\
  \bibinfo {author} {\bibfnamefont {B.}~\bibnamefont {Shao}},\ }\bibfield
  {title} {\bibinfo {title} {{Nondemolition quasiprobabilities of work and heat
  in the presence of a non-Markovian environment}},\ }\href
  {https://doi.org/10.1103/PhysRevA.109.032228} {\bibfield  {journal} {\bibinfo
   {journal} {Phys. Rev. A}\ }\textbf {\bibinfo {volume} {109}},\ \bibinfo
  {pages} {032228} (\bibinfo {year} {2024})}\BibitemShut {NoStop}%
\bibitem [{\citenamefont {Lalita}\ and\ \citenamefont
  {Banerjee}(2025)}]{Lalita2025}%
  \BibitemOpen
  \bibfield  {author} {\bibinfo {author} {\bibfnamefont {J.}~\bibnamefont
  {Lalita}}\ and\ \bibinfo {author} {\bibfnamefont {S.}~\bibnamefont
  {Banerjee}},\ }\bibfield  {title} {\bibinfo {title} {{Non-classicality of
  two-qubit quantum collision model: non-Markovian effects}},\ }\href@noop {}
  {\bibfield  {journal} {\bibinfo  {journal} {arXiv preprint arXiv:2506.23818}\
  } (\bibinfo {year} {2025})}\BibitemShut {NoStop}%
\bibitem [{\citenamefont {Luo}\ \emph {et~al.}(2012{\natexlab{b}})\citenamefont
  {Luo}, \citenamefont {Fu},\ and\ \citenamefont {Song}}]{PhysRevA.86.044101}%
  \BibitemOpen
  \bibfield  {author} {\bibinfo {author} {\bibfnamefont {S.}~\bibnamefont
  {Luo}}, \bibinfo {author} {\bibfnamefont {S.}~\bibnamefont {Fu}},\ and\
  \bibinfo {author} {\bibfnamefont {H.}~\bibnamefont {Song}},\ }\bibfield
  {title} {\bibinfo {title} {Quantifying non-markovianity via correlations},\
  }\href {https://doi.org/10.1103/PhysRevA.86.044101} {\bibfield  {journal}
  {\bibinfo  {journal} {Phys. Rev. A}\ }\textbf {\bibinfo {volume} {86}},\
  \bibinfo {pages} {044101} (\bibinfo {year} {2012}{\natexlab{b}})}\BibitemShut
  {NoStop}%
\bibitem [{\citenamefont {Solinas}\ and\ \citenamefont
  {Gherardini}(2025)}]{solinas2024quantum}%
  \BibitemOpen
  \bibfield  {author} {\bibinfo {author} {\bibfnamefont {P.}~\bibnamefont
  {Solinas}}\ and\ \bibinfo {author} {\bibfnamefont {S.}~\bibnamefont
  {Gherardini}},\ }\bibfield  {title} {\bibinfo {title} {Negativity of
  nondemolition quasiprobability distribution as a necessary and sufficient
  condition for macrorealism violation},\ }\href
  {https://doi.org/10.1103/PhysRevA.111.052217} {\bibfield  {journal} {\bibinfo
   {journal} {Phys. Rev. A}\ }\textbf {\bibinfo {volume} {111}},\ \bibinfo
  {pages} {052217} (\bibinfo {year} {2025})}\BibitemShut {NoStop}%
\bibitem [{\citenamefont {Melegari}\ \emph {et~al.}(2025)\citenamefont
  {Melegari}, \citenamefont {Cardi},\ and\ \citenamefont
  {Solinas}}]{melegari2025}%
  \BibitemOpen
  \bibfield  {author} {\bibinfo {author} {\bibfnamefont {D.}~\bibnamefont
  {Melegari}}, \bibinfo {author} {\bibfnamefont {M.}~\bibnamefont {Cardi}},\
  and\ \bibinfo {author} {\bibfnamefont {P.}~\bibnamefont {Solinas}},\
  }\bibfield  {title} {\bibinfo {title} {Quantum simulations of macrorealism
  violation via the quantum nondemolition measurement protocol},\ }\href
  {https://doi.org/10.1103/PhysRevA.111.052435} {\bibfield  {journal} {\bibinfo
   {journal} {Phys. Rev. A}\ }\textbf {\bibinfo {volume} {111}},\ \bibinfo
  {pages} {052435} (\bibinfo {year} {2025})}\BibitemShut {NoStop}%
\bibitem [{\citenamefont {Solinas}\ and\ \citenamefont
  {Gherardini}(2026)}]{SolinasReview2026}%
  \BibitemOpen
  \bibfield  {author} {\bibinfo {author} {\bibfnamefont {P.}~\bibnamefont
  {Solinas}}\ and\ \bibinfo {author} {\bibfnamefont {S.}~\bibnamefont
  {Gherardini}},\ }\bibfield  {title} {\bibinfo {title} {Quantumness
  certification via non-demolition measurements},\ }\href
  {https://doi.org/https://doi.org/10.1080/23746149.2026.2637832} {\bibfield
  {journal} {\bibinfo  {journal} {Advances in Physics: X}\ }\textbf {\bibinfo
  {volume} {11}},\ \bibinfo {pages} {2637832} (\bibinfo {year}
  {2026})}\BibitemShut {NoStop}%
\bibitem [{\citenamefont {Wang}\ \emph {et~al.}(2024)\citenamefont {Wang},
  \citenamefont {Wu}, \citenamefont {Yao}, \citenamefont {Lian}, \citenamefont
  {Cheng}, \citenamefont {Xu}, \citenamefont {Zhang}, \citenamefont {Jiang},
  \citenamefont {Xu}, \citenamefont {Qi}, \citenamefont {Hou}, \citenamefont
  {Zhou}, \citenamefont {He},\ and\ \citenamefont {Duan}}]{WangPRA2024}%
  \BibitemOpen
  \bibfield  {author} {\bibinfo {author} {\bibfnamefont {G.-X.}\ \bibnamefont
  {Wang}}, \bibinfo {author} {\bibfnamefont {Y.-K.}\ \bibnamefont {Wu}},
  \bibinfo {author} {\bibfnamefont {R.}~\bibnamefont {Yao}}, \bibinfo {author}
  {\bibfnamefont {W.-Q.}\ \bibnamefont {Lian}}, \bibinfo {author}
  {\bibfnamefont {Z.-J.}\ \bibnamefont {Cheng}}, \bibinfo {author}
  {\bibfnamefont {Y.-L.}\ \bibnamefont {Xu}}, \bibinfo {author} {\bibfnamefont
  {C.}~\bibnamefont {Zhang}}, \bibinfo {author} {\bibfnamefont
  {Y.}~\bibnamefont {Jiang}}, \bibinfo {author} {\bibfnamefont {Y.-Z.}\
  \bibnamefont {Xu}}, \bibinfo {author} {\bibfnamefont {B.-X.}\ \bibnamefont
  {Qi}}, \bibinfo {author} {\bibfnamefont {P.-Y.}\ \bibnamefont {Hou}},
  \bibinfo {author} {\bibfnamefont {Z.-C.}\ \bibnamefont {Zhou}}, \bibinfo
  {author} {\bibfnamefont {L.}~\bibnamefont {He}},\ and\ \bibinfo {author}
  {\bibfnamefont {L.-M.}\ \bibnamefont {Duan}},\ }\bibfield  {title} {\bibinfo
  {title} {Simulating the spin-boson model with a controllable reservoir in an
  ion trap},\ }\href {https://doi.org/10.1103/PhysRevA.109.062402} {\bibfield
  {journal} {\bibinfo  {journal} {Phys. Rev. A}\ }\textbf {\bibinfo {volume}
  {109}},\ \bibinfo {pages} {062402} (\bibinfo {year} {2024})}\BibitemShut
  {NoStop}%
\bibitem [{\citenamefont {Sun}\ \emph {et~al.}(2025)\citenamefont {Sun},
  \citenamefont {Kang}, \citenamefont {Nuomin}, \citenamefont {Schwartz},
  \citenamefont {Beratan}, \citenamefont {Brown},\ and\ \citenamefont
  {Kim}}]{SunNatComm2025}%
  \BibitemOpen
  \bibfield  {author} {\bibinfo {author} {\bibfnamefont {K.}~\bibnamefont
  {Sun}}, \bibinfo {author} {\bibfnamefont {M.}~\bibnamefont {Kang}}, \bibinfo
  {author} {\bibfnamefont {H.}~\bibnamefont {Nuomin}}, \bibinfo {author}
  {\bibfnamefont {G.}~\bibnamefont {Schwartz}}, \bibinfo {author}
  {\bibfnamefont {D.~N.}\ \bibnamefont {Beratan}}, \bibinfo {author}
  {\bibfnamefont {K.~R.}\ \bibnamefont {Brown}},\ and\ \bibinfo {author}
  {\bibfnamefont {J.}~\bibnamefont {Kim}},\ }\bibfield  {title} {\bibinfo
  {title} {Quantum simulation of spin-boson models with structured bath},\
  }\href {https://doi.org/10.1038/s41467-025-59296-y} {\bibfield  {journal}
  {\bibinfo  {journal} {Nat. Commun.}\ }\textbf {\bibinfo {volume} {16}},\
  \bibinfo {pages} {4042} (\bibinfo {year} {2025})}\BibitemShut {NoStop}%
\bibitem [{\citenamefont {Haase}\ \emph {et~al.}(2018)\citenamefont {Haase},
  \citenamefont {Vetter}, \citenamefont {Unden}, \citenamefont {Smirne},
  \citenamefont {Rosskopf}, \citenamefont {Naydenov}, \citenamefont {Stacey},
  \citenamefont {Jelezko}, \citenamefont {Plenio},\ and\ \citenamefont
  {Huelga}}]{HaasePRL2018}%
  \BibitemOpen
  \bibfield  {author} {\bibinfo {author} {\bibfnamefont {J.~F.}\ \bibnamefont
  {Haase}}, \bibinfo {author} {\bibfnamefont {P.~J.}\ \bibnamefont {Vetter}},
  \bibinfo {author} {\bibfnamefont {T.}~\bibnamefont {Unden}}, \bibinfo
  {author} {\bibfnamefont {A.}~\bibnamefont {Smirne}}, \bibinfo {author}
  {\bibfnamefont {J.}~\bibnamefont {Rosskopf}}, \bibinfo {author}
  {\bibfnamefont {B.}~\bibnamefont {Naydenov}}, \bibinfo {author}
  {\bibfnamefont {A.}~\bibnamefont {Stacey}}, \bibinfo {author} {\bibfnamefont
  {F.}~\bibnamefont {Jelezko}}, \bibinfo {author} {\bibfnamefont {M.~B.}\
  \bibnamefont {Plenio}},\ and\ \bibinfo {author} {\bibfnamefont {S.~F.}\
  \bibnamefont {Huelga}},\ }\bibfield  {title} {\bibinfo {title} {{Controllable
  Non-Markovianity for a Spin Qubit in Diamond}},\ }\href
  {https://doi.org/10.1103/PhysRevLett.121.060401} {\bibfield  {journal}
  {\bibinfo  {journal} {Phys. Rev. Lett.}\ }\textbf {\bibinfo {volume} {121}},\
  \bibinfo {pages} {060401} (\bibinfo {year} {2018})}\BibitemShut {NoStop}%
\bibitem [{\citenamefont {Dalacu}\ \emph {et~al.}(2019)\citenamefont {Dalacu},
  \citenamefont {Poole},\ and\ \citenamefont {Williams}}]{Dalacu_2019}%
  \BibitemOpen
  \bibfield  {author} {\bibinfo {author} {\bibfnamefont {D.}~\bibnamefont
  {Dalacu}}, \bibinfo {author} {\bibfnamefont {P.~J.}\ \bibnamefont {Poole}},\
  and\ \bibinfo {author} {\bibfnamefont {R.~L.}\ \bibnamefont {Williams}},\
  }\bibfield  {title} {\bibinfo {title} {Nanowire-based sources of
  non-classical light},\ }\href {https://doi.org/10.1088/1361-6528/ab0393}
  {\bibfield  {journal} {\bibinfo  {journal} {Nanotechnology}\ }\textbf
  {\bibinfo {volume} {30}},\ \bibinfo {pages} {232001} (\bibinfo {year}
  {2019})}\BibitemShut {NoStop}%
\bibitem [{\citenamefont {Sapienza}\ \emph {et~al.}(2015)\citenamefont
  {Sapienza}, \citenamefont {Davan{\c{c}}o}, \citenamefont {Badolato},\ and\
  \citenamefont {Srinivasan}}]{Sapienza2015}%
  \BibitemOpen
  \bibfield  {author} {\bibinfo {author} {\bibfnamefont {L.}~\bibnamefont
  {Sapienza}}, \bibinfo {author} {\bibfnamefont {M.}~\bibnamefont
  {Davan{\c{c}}o}}, \bibinfo {author} {\bibfnamefont {A.}~\bibnamefont
  {Badolato}},\ and\ \bibinfo {author} {\bibfnamefont {K.}~\bibnamefont
  {Srinivasan}},\ }\bibfield  {title} {\bibinfo {title} {Nanoscale optical
  positioning of single quantum dots for bright and pure single-photon
  emission},\ }\href {https://doi.org/10.1038/ncomms8833} {\bibfield  {journal}
  {\bibinfo  {journal} {Nature Communications}\ }\textbf {\bibinfo {volume}
  {6}},\ \bibinfo {pages} {7833} (\bibinfo {year} {2015})}\BibitemShut
  {NoStop}%
\bibitem [{\citenamefont {Mirhosseini}\ \emph {et~al.}(2019)\citenamefont
  {Mirhosseini}, \citenamefont {Kim}, \citenamefont {Zhang}, \citenamefont
  {Sipahigil}, \citenamefont {Dieterle}, \citenamefont {Keller}, \citenamefont
  {Asenjo-Garcia}, \citenamefont {Chang},\ and\ \citenamefont
  {Painter}}]{Mirhosseini2019}%
  \BibitemOpen
  \bibfield  {author} {\bibinfo {author} {\bibfnamefont {M.}~\bibnamefont
  {Mirhosseini}}, \bibinfo {author} {\bibfnamefont {E.}~\bibnamefont {Kim}},
  \bibinfo {author} {\bibfnamefont {X.}~\bibnamefont {Zhang}}, \bibinfo
  {author} {\bibfnamefont {A.}~\bibnamefont {Sipahigil}}, \bibinfo {author}
  {\bibfnamefont {P.~B.}\ \bibnamefont {Dieterle}}, \bibinfo {author}
  {\bibfnamefont {A.~J.}\ \bibnamefont {Keller}}, \bibinfo {author}
  {\bibfnamefont {A.}~\bibnamefont {Asenjo-Garcia}}, \bibinfo {author}
  {\bibfnamefont {D.~E.}\ \bibnamefont {Chang}},\ and\ \bibinfo {author}
  {\bibfnamefont {O.}~\bibnamefont {Painter}},\ }\bibfield  {title} {\bibinfo
  {title} {Cavity quantum electrodynamics with atom-like mirrors},\ }\href
  {https://doi.org/10.1038/s41586-019-1196-1} {\bibfield  {journal} {\bibinfo
  {journal} {Nature}\ }\textbf {\bibinfo {volume} {569}},\ \bibinfo {pages}
  {692} (\bibinfo {year} {2019})}\BibitemShut {NoStop}%
\end{thebibliography}%

\appendix*
\section{APPENDIX}

\subsection{Iteration procedure of the system's open dynamics}

Starting from the joint state $\rho_{SM}^{(n-1)}$ of the system and memory at step $n-1$, and from the ``fresh'' environment particle initialized in the default state $\rho_A^{(n)}$, we have:
\begin{eqnarray}
\bar{\rho}_{SMA}^{(n)} &=& U_{SM} \big( \rho^{(n-1)}_{SM} \otimes \rho_A \big) U^{\dagger}_{SM} \equiv \bar{\rho}_{SM}^{(n)} \otimes \rho_A,
\label{eq:collisionmodel_SM}
\\
\rho_{SMA}^{(n)} &=& U_{MA}^{(n)} \big( \bar{\rho}_{SM}^{(n)} \otimes \rho_A \big) U^{(n)\,\dagger}_{MA},
\label{eq:collisionmodel_MA}
\end{eqnarray}
where $\bar{\rho}_{SM}^{(n)} = \text{Tr}_{A}\left[ \bar{\rho}_{SMA}^{(n)} \right]$ is the intermediate state of the system and memory, after the $S$-$M$ collision but before the $M$-$A$ collision. $\rho^{(n)}_{SM} = \text{Tr}_{A}\left[ \rho_{SMA}^{(n)} \right]$ is the updated state after both interactions, and after the environment particle has been traced out. The state $\rho^{(n)}_{SM}$ is carried over to the next iteration.

\subsection{Phase covariant quantum maps}

Let us consider a collision model generated by local Hamiltonians diagonal in the $\sigma_z$-basis and by Heisenberg-type interactions conserving total $z$-magnetization. Thus, the reduced dynamics of the system is covariant with respect to local $z$-rotations~\cite{filippov2020,Smirne2016,siudzinska2023,Holevo1993}. This means that for every angle $\phi$, taking $R_\phi = e^{-i\phi\sigma_z/2}$, we have that
\begin{equation}
\label{eq:covariance}
\Lambda_{n,n-1}\left[ R_\phi \rho R_\phi^\dagger \right] = R_{\phi}\,\Lambda_{n,n-1}[\rho]\,R_\phi^{\dagger}.
\end{equation}

In the matrix-elements basis for operators on $S$, that is $\ket{\rho_S}\!\rangle = (\rho_{00},\rho_{01},\rho_{10},\rho_{11})^T$, the action of the rotation superoperator is diagonal:
\begin{equation}
R_\phi \rho R_\phi^\dagger
\;\longleftrightarrow\;
\begin{pmatrix}
1 & 0 & 0 & 0\\
0 & e^{-i\phi} & 0 & 0\\
0 & 0 & e^{i\phi} & 0\\
0 & 0 & 0 & 1
\end{pmatrix}
\begin{pmatrix}
\rho_{00}\\
\rho_{01}\\
\rho_{10}\\
\rho_{11}
\end{pmatrix}.
\end{equation}
Covariance implies that the population terms $\rho_{00}$ and $\rho_{11}$ can mix only with each other, while the quantum coherence $\rho_{01}$ and $\rho_{10}$ evolve independently and do not couple to populations. Therefore, the most general intermediate map compatible with phase covariance has the form
\begin{equation}
\label{eq:ansatz_general}
\Lambda_{n,n-1} \equiv \Lambda=
\begin{pmatrix}
a & 0 & 0 & b\\
0 & c & d & 0\\
0 & e & f & 0\\
g & 0 & 0 & h
\end{pmatrix}.
\end{equation}
If the map preserves Hermiticity, then necessarily $f=c^*$ and $e=d^*$. Moreover, if the map is trace-preserving, then
$a+g=1$ and $b+h=1$. Accordingly, the most general Hermiticity-preserving and trace-preserving phase-covariant qubit map is
\begin{equation}
\label{eq:ansatz_app}
\Lambda =
\begin{pmatrix}
a & 0 & 0 & b\\
0 & c & d & 0\\
0 & d^* & c^* & 0\\
1-a & 0 & 0 & 1-b
\end{pmatrix}.
\end{equation}
Equivalently, its action on density matrices is
\begin{equation}
\label{eq:map_action}
\Lambda[\rho]=
\begin{pmatrix}
a\,\rho_{00}+b\,\rho_{11} & c\,\rho_{01}+d\,\rho_{10}\\[1mm]
c^*\,\rho_{10}+d^*\,\rho_{01} & (1-a)\rho_{00}+(1-b)\rho_{11}
\end{pmatrix}.
\end{equation}
Notice that $d \neq 0$ when excitation number is not conserved due to anisotropic couplings. $d$ is still vanishing in the non-energy preserving regime due to a non-zero system-memory detuning: $\Delta\omega = \omega_S - \omega_M \neq 0$.

\subsection{Quantum process tomography}

We here discuss the process tomography adopted to numerically reconstruct the dynamical map $\Lambda_n \equiv \Lambda_{n,0}$ for any $n > 0$. Once the tomography is completed, we are able to compute evolved states at any time from any input state, without the need to actually run the input state through the original dynamical process. In practice, for the reconstruction, we need to run the dynamical map on 4 different input states and store the whole history of their time evolution.    

Once the $\Lambda_n$ has been reconstructed, we can compute its inverse $\Lambda_n^{-1}$ and the time local maps $\Lambda_{n,n-1} = \Lambda_n \circ \Lambda_{n-1}^{-1}$ for $n>0$, which will allow us then to estimate the LFS measure of non-Markovianity.  

\subsubsection{General map}

Throughout the section we will employ the Bloch vector representation for density matrices:
\begin{equation}\label{eq:def_Bloch_representation}
    \rho_S^{(n)} = \frac{\mathbb{I} + r^{(n)}_x\sigma_x + r^{(n)}_y\sigma_y + r^{(n)}_z \sigma_z }{2}
\end{equation}
with $\vec{r}_n = \left( r^{(n)}_x,r^{(n)}_y,r^{(n)}_z \right)^T$, where $\mathbb{I}$ is the $2 \times 2$ identity matrix and $\sigma_{x,y,z}$ are the Pauli matrices. The dynamical process can be expressed as a family of transformations, for all steps $n>0$, from initial to final Bloch vectors:
\begin{equation}
    \rho_S^{(0)} \mapsto \rho_S^{(n)} = \Lambda_n[\rho_S^{(0)}] \quad \Longleftrightarrow \quad \vec{r}_0 \mapsto \vec{r}_n.
\end{equation}
The problem of reconstructing $\Lambda$ is formulated in terms of finding this family of transformations. 
Given the linearity of quantum mechanics, without loss of generality, the transformation $\vec{r}_0 \mapsto \vec{r}_n$ can be taken as an affine transformation within the space of Bloch vectors:
\begin{equation}\label{eq:BlochVectorMap}
    \vec{r}_n = M_n \vec{r}_0 + \vec{c}_n \,, 
\end{equation}
where $\{ M_n \}$ for $n \geq 0$ is a family of $3 \times 3$ matrices and $\{ \vec{c}_n \}$ denotes a family of constant shift vectors.

Strictly speaking, the map $\Lambda$ is supposed to send density matrices into density matrices, and it should not be applied to objects which are not proper density matrices. However, let us forget about this for a moment and treat $\Lambda$ as a mere computational tool that we can apply linearly on any $2 \times 2$ matrix, irrespectively of whether it is Hermitian or not, or whether it is a unit-trace matrix or not; we will come back to this point shortly. Hence, we express the final states by making use of the Bloch representation: 
\begin{equation}
\label{eq:evolvedRho}
    \rho_S^{(n)} = \Lambda_n[\rho_S^{(0)}] = \frac{\Lambda_n[\mathbb{I}] + r^{(0)}_x \Lambda_n[\sigma_x] + r^{(0)}_y \Lambda_n[\sigma_y] + r^{(0)}_z \Lambda_n[\sigma_z]}{2}.
\end{equation}
Comparing \eqref{eq:evolvedRho} with the Bloch representation of the evolved state [Eq.~\eqref{eq:BlochVectorMap}], we have that
\begin{equation}
\label{eq:cvec}
    \vec{c}_n = 
    \begin{pmatrix}
    c^{(n)}_x \\
    c^{(n)}_y \\
    c^{(n)}_z
    \end{pmatrix}
    = \frac{1}{2}
    \begin{pmatrix}
    \text{Tr}\left[ \sigma_x \Lambda_n[\mathbb{I}] \right] \\
    \text{Tr} \left[ \sigma_y \Lambda_n[\mathbb{I}] \right] \\
    \text{Tr}\left[ \sigma_z \Lambda_n[\mathbb{I}] \right]
    \end{pmatrix}.
\end{equation}
Similarly, by direct substitution, one has that
\begin{equation}
\label{eq:Mmat}
    [M_n]_{i,j} = \frac{1}{2} \text{Tr} \left[ \sigma_i \Lambda_n [\sigma_j] \right], \quad i,j=x,y,z.
\end{equation}
Therefore, if we determine the action of $\Lambda_n$ on the identity and on the three Pauli matrices, for any $n$, we can compute any evolved state from any arbitrary input state using Eq.~\eqref{eq:evolvedRho}.

Let us address now the issue mentioned above, namely that $\Lambda$ should only be applied to proper density matrices. We use the usual representation where the basis vectors are the eigenstates of $\sigma_z$: $\sigma_z \ket{0} = \ket{0}$ and $\sigma_z \ket{1} = -\ket{1}$. Let us also recall the eigenstates of the other two Pauli matrices, $\sigma_x \ket{\pm} = \pm \ket{\pm}$, with $\ket{\pm} = (\ket{0} \pm \ket{1})/\sqrt{2}$, and  $\sigma_y \ket{R} = +\ket{R}$, $\sigma_y \ket{L} = - \ket{L}$, with $\ket{R,L} = (\ket{0} \pm i \ket{1})/\sqrt{2}$. Then, we consider the following projectors on some of these eigenstates:
\begin{eqnarray}
&&P_0 = \ketbra{0}{0} = 
\begin{pmatrix}
1 & 0 \\
0 & 0
\end{pmatrix},
\qquad\qquad\quad\,\,
P_1 = \ketbra{1}{1} = 
\begin{pmatrix}
0 & 0 \\
0 & 1
\end{pmatrix},\nonumber\\
&&P_+ = \ketbra{+}{+} = \frac{1}{2} 
\begin{pmatrix}
1 & 1 \\
1 & 1
\end{pmatrix}, 
\qquad\qquad
P_R = \ketbra{R}{R} = \frac{1}{2}
\begin{pmatrix}
1 & -i \\
i & 1
\end{pmatrix},\nonumber\\
\end{eqnarray}
from which one ends up to the following identities:
\begin{eqnarray}
\label{eq:PauliProjectors}
&&\mathbb{I} = P_0 + P_1, 
\qquad\qquad\qquad\qquad
\sigma_x = 2 P_{+} - (P_0 + P_1), 
\nonumber\\
&&\sigma_y = 2 P_{R} - (P_0 + P_1), 
\qquad\qquad
\sigma_z = P_0 - P_1.
\end{eqnarray}
Thanks to the linearity of the map, we can express $\Lambda_n[\mathbb{I}]$, $\Lambda_n[\sigma_x]$, $\Lambda_n[\sigma_y]$ and $\Lambda_n[\sigma_z]$ in terms of $\Lambda_n[P_0]$, $\Lambda_n[P_1]$, $\Lambda_n[P_+]$ and $\Lambda_n[P_R]$ using relations that are perfectly similar to those in Eq.~\eqref{eq:PauliProjectors}. Note that now the use of $\Lambda$ is legitimate, since the four projectors $P_0, P_1, P_+, P_R$ are proper density matrices. Therefore, we choose these four projectors as input ``probe'' states for the tomography; indeed, by reconstructing the time evolution of the four probe states, we can compute the evolved states at any time from a generic input state, as desired.

To summarize, the procedure to perform the tomography of $\Lambda$ is as follows:
\begin{enumerate}
    \item 
    Run the dynamics of the system initializing it in the four probe states $P_0, P_1, P_+, P_R$, in order to obtain their whole time evolutions $\{ \Lambda_n[P_0] \}$, $\{ \Lambda_n[P_1] \}$, $\{ \Lambda_n[P_+] \}$, $\{ \Lambda_n[P_R] \}$ for all $n$.    
    \item 
    At each step $n$, from $\Lambda_n[P_0]$, $\Lambda_n[P_1]$, $\Lambda_n[P_+]$, $\Lambda_n[P_R]$ compute $\Lambda_n[\mathbb{I}]$, $\Lambda_n[\sigma_x]$, $\Lambda_n[\sigma_y]$, $\Lambda_n[\sigma_z]$ through Eq.~\eqref{eq:PauliProjectors}, using the linearity of $\Lambda$.
    \item 
    Construct the vectors $\{ \vec{c}_n \}$, Eq.~\eqref{eq:cvec} and matrices $\{ M_n \}$, Eq.~\eqref{eq:Mmat} for all $n$.
\end{enumerate}
In this way, we can compute the evolved state via Eqs.~\eqref{eq:def_Bloch_representation}-\eqref{eq:BlochVectorMap}, initializing the system in an arbitrary input state.    

\subsubsection{Inverted map}

Assuming $\Lambda$ has been successfully reconstructed, we show how to derive $\Lambda^{-1}$. As in \eqref{eq:BlochVectorMap}, we can express the action of $\Lambda_n^{-1}$ on a generic (evolved) Bloch vector:
\begin{equation}
\label{eq:InvertedBlochVectorMap}
\rho_S^{(0)} = \Lambda_n^{-1}\left[ \rho_S^{(n)} \right] \quad \longrightarrow \quad  \vec{r}_0 = \widetilde{M}_n \vec{r}_n + \vec{\widetilde{c}}_n. 
\end{equation}    
To obtain $\widetilde{M}_n$ and $\vec{\widetilde{c}}_n$, we express the identity $\rho_S^{(0)} = \Lambda_n^{-1}\circ\Lambda_n\left[ \rho_S^{(0)} \right]$ in the Bloch vector form using Eqs.~\eqref{eq:BlochVectorMap}-\eqref{eq:InvertedBlochVectorMap}:
\begin{equation}\label{eq:vec-r_0}
    \vec{r}_0 = \widetilde{M}_n\left( M_n\vec{r}_0 + \vec{c}_n \right) + \vec{\widetilde{c}}_n.
\end{equation}
Requesting \eqref{eq:vec-r_0} to be valid for any $\vec{r}_0$ leads to $\widetilde{M}_{n}M_n = \mathbb{I}$, namely $\widetilde{M}_n = M_n^{-1}$. Then, by taking $\vec{r}_0 = \vec{0}$, we find the expression of $\vec{\widetilde{c}}_n$. We get: $\widetilde{M}_n = M_n^{-1}$ and $\vec{\widetilde{c}}_n = -M_n^{-1}\vec{c}_n$. Notice that we are able to reconstruct $\Lambda^{-1}$ if the matrices $\{ M_n \}$ are invertible, that is, they are full rank matrices for all $n$. 

\subsubsection{Time-local map}

We have now all the ingredients for the reconstruction of $\Lambda_{n,n-1} = \Lambda_n \circ \Lambda_{n-1}^{-1}$ that, as explained in the main text, is sufficient to compute any arbitrary-time map $\Lambda_{n,m}$. 
Let us thus express the action of $\Lambda_{n,n-1}$ with Bloch vectors:
\begin{equation}
\label{eq:BlochVectorMapD}
\rho_S^{(n)} = \Lambda_{n,n-1}\left[ \rho_S^{(n-1)} \right] \quad \longrightarrow \quad
\vec{r}_n = M^{\text{tl}}_n \, \vec{r}_{n-1} + \vec{c^{\text{tl}}_n}, 
\end{equation}
where the superscript ``tl'' stands for ``time-local''. Having reconstructed $\Lambda_n$ and $\Lambda_n^{-1}$ for all $n$, the action of $\Lambda_{n,n-1}\left[ \rho_S^{(n-1)} \right] = \Lambda_n \circ \Lambda_{n-1}^{-1}\left[ \rho_S^{(n-1)}\right]$ is expressed in the formalism of Bloch vectors using Eqs.~\eqref{eq:BlochVectorMap}-\eqref{eq:InvertedBlochVectorMap}: 
\begin{equation}\label{eq:vec-r_n}
    \vec{r}_n = M_n\left( \widetilde{M}_{n-1}\vec{r}_{n-1} + \vec{\widetilde{c}}_{n-1}\right) + \vec{c}_n. 
\end{equation}
By comparing Eq.~\eqref{eq:vec-r_n} with Eq.~\eqref{eq:BlochVectorMapD}, we obtain that $M^{\text{tl}}_n = M_{n}M^{-1}_{n-1}$ and $\vec{c^{\text{tl}}_n} = M_n\vec{\widetilde{c}}_{n-1} + \vec{c}_n$. This completes the reconstruction of the family of maps $\Lambda_{n,n-1}$ for all $n$.

\subsection{RHP measure of non-Markovianity}

The RHP measure of non-Markovianity for quantum dynamics relies on violations of CP-divisibility~\cite{RHP-non-Markovianity}. We are going to briefly recall it for a generic discrete-time process.

RHP measure requires full knowledge of the family of maps $\{ \Lambda_n \}$, with $\Lambda_n \equiv \Lambda_{n,0}$, for any $n \geq 0$. Each map $\Lambda_n$ is assumed to be invertible such that $\Lambda_n^{-1}$ can be computed. The assumption that the maps $\{\Lambda_n\}$ are invertible usually does not represent a restriction. Unitary channels and all dynamical maps resulting from integrating a master equation satisfy it. This assumption would represent a problem only for processes that alter the rank of the system density matrix, such as measurement processes. Hence, an arbitrary 2-time map is obtained as $\Lambda_{t,s} \equiv \Lambda_t \circ \Lambda_s^{-1}$, for all $0<s<t$. The RHP measure affirms: quantum dynamics are \emph{Markovian} if the corresponding map is always \emph{CP-divisible}, that is, if the maps $\Lambda_{t,s}$ are CP for all $0<s<t$. The dynamics are \emph{non-Markovian} if, for some $s,t$, the map $\Lambda_{t,s}$ is not CP; in such a case, the whole dynamics is said to be non-CP-divisible.

A local CP trace-preserving channel can never increase the amount of entanglement between the open system $S$ and another identical copy $R$ of itself (reference system); this is the case for Markovian maps~\cite{RHP-non-Markovianity}. This rationale translates into the RHP measure through the Choi-Jamio\l{}kowski isomorphism, which is used to determine whether the map $\Lambda_{t,s}$ is CP. Formally, given a maximally entangled state $\ket{\Psi}$ of $S$ and $R$, the map $\Lambda_{t,s}$ is CP if and only if $(\Lambda_{t,s} \otimes \mathbb{I})[\ketbra{\Psi}]$ is positive semi-definite. Hence, the quantity 
\begin{equation}
f_{\text{NCP}}(t,s) \equiv \Big\Vert (\Lambda_{t,s} \otimes \mathbb{I})\left[ \ketbra{\Psi}{\Psi} \right] \Big\Vert_1
\end{equation}
is a measure of the non-CP character of $\Lambda_{t,s}$: $f_{\text{NCP}}(t,s) = 1$ if $\Lambda_{t,s}$ is CP, and $f_{\text{NCP}}(t,s) > 1$ otherwise. In particular,
\begin{equation}\label{eq:g}
    g_n = \Big\Vert (\Lambda_{n,n-1} \otimes \mathbb{I}) \big[ \ketbra{\Psi}{\Psi} \big] \Big\Vert_1 - 1
\end{equation}
is strictly $0$ if $\Lambda_{t,s}$ is CP $\forall \, 0<s<t$, and greater than 0 otherwise. Overall, the non-Markovianity measure based on violations of CP-divisibility is 
\begin{equation}
\mathcal{I_{\rm RHP}} = \sum_{n=1}^{n_{\text{max}}} g_{n}.
\end{equation}

\subsection{LFS measure of non-Markovianity}

The total correlations of a bipartite state $\rho_{LS}$, with $S$ the system under scrutiny and $L$ an auxiliary system, is quantified by the QMI $I(\rho_{LS}) = S(\rho_L) + S(\rho_S) - S(\rho_{LS})$ with $S(\rho) = -\rm{Tr}[\rho \log_2(\rho)]$. If the map $\Lambda_n$ is Markovian, then the QMI evolves as 
\begin{equation}
I(\rho_{LS}^{(n)}) = I\left( (\mathbb{I}_L \otimes \Lambda_{n,s} \Lambda_s) [\rho_{LS}^{(0)}] \right) = I\left( (\mathbb{I}_L \otimes \Lambda_{n,s})[\rho_{LS}^{(s)}] \right) \leq I(\rho_{LS}^{(s)})
\end{equation}
with $s \le n$, due to the monotonicity of the QMI under local operations~\cite{PhysRevA.86.044101}. As such, we have $\Delta I_{LS}^{(n)} \equiv I(\rho_{LS}^{(n+1)})-I(\rho_{LS}^{(n)})\leq 0$. Any violation of the latter inequality is a signature of non-Markovianity and is measured via
\begin{equation}\label{eq:def_I_LFS}
    \mathcal{I}_{\text{LFS}} = \sup_{\rho_{LS}^{(0)}} \sum_{k\,\text{s.t.}\,\Delta I_{LS}^{(k)}>0}\Delta I_{LS}^{(k)}.
\end{equation}
Initializing $S, L$ in any maximally correlated pure state $\rho_{LS}^{(0)}$, wherein $S, L$ occupy Hilbert spaces of equal dimension, \eqref{eq:def_I_LFS} reduces to 
\begin{equation}
\mathcal{I}_{\text{LFS}} = \sum_{k\,\text{s.t.}\,\Delta I_{LS}^{(k)}>0}\Delta I_{LS}^{(k)},
\end{equation}
thus removing the necessity for optimization over all possible initial states. Notice that, for initial states $\rho_{LS}^{(0)}$ that are not maximally correlated, $\mathcal{I}_{\text{LFS}} = \sum_{k\,\text{s.t.}\,\Delta I_{LS}^{(k)}>0}\Delta I_{LS}^{(k)}$ corresponds to a lower bound (thus, a witness) of non-Markovianity, rather than a measure.

\begin{figure}[t]
    \centering
    \includegraphics[width=0.9\columnwidth]{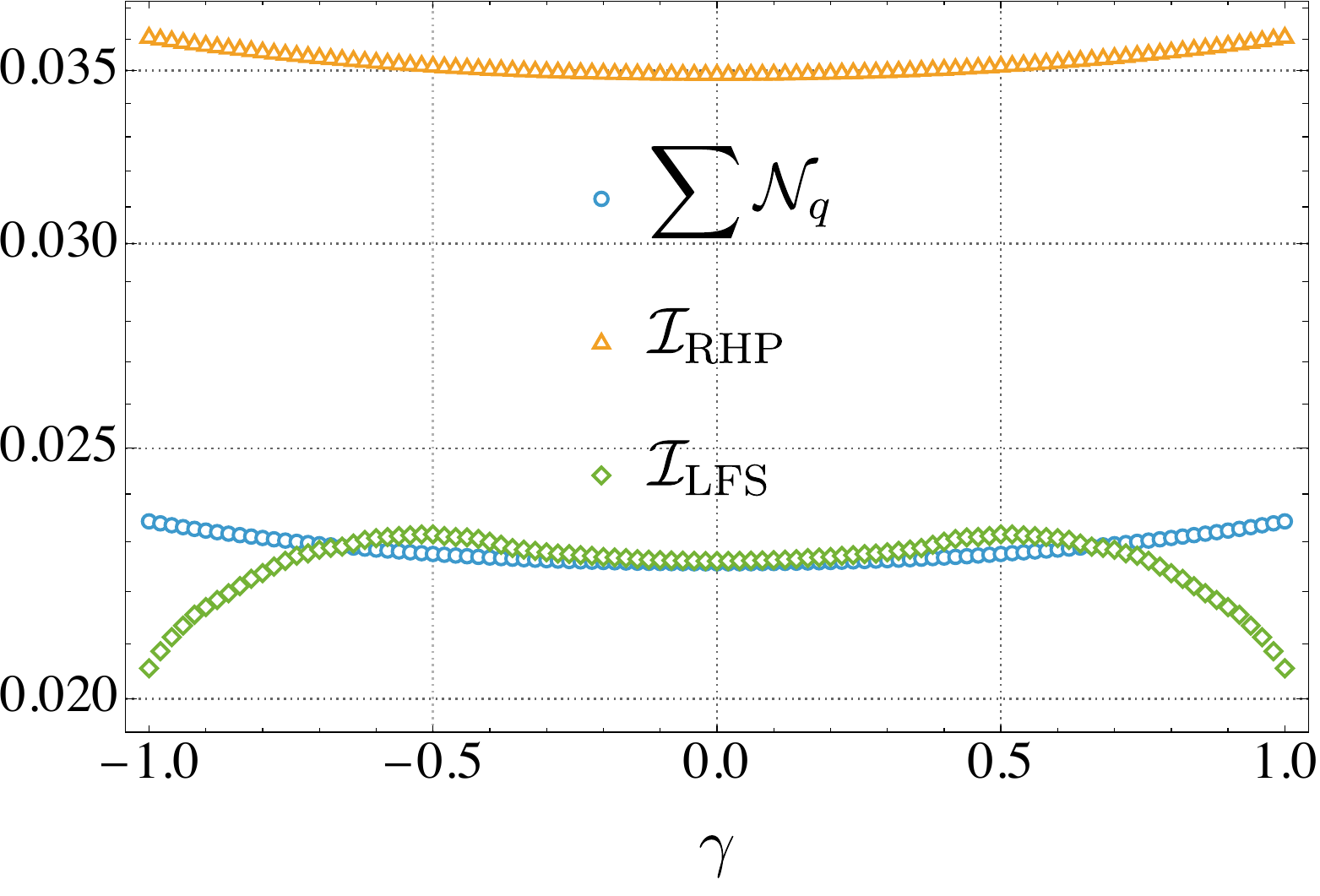}
    \caption{
    Comparison of $\mathcal{I}_{\text{LFS}}$ (green rhombs), $\mathcal{I}_{\text{RHP}}$ (orange triangles) and $\sum_{n}\mathcal N_{\mathrm q}[P_n]$ (light blue circles), as a function of the coupling anisotropy $\gamma\in[-1,1]$ for system-memory collisions. The interaction between memory and environment particles is kept energy-preserving. Moreover, the maximum number of collisions is fixed to $10^3$. All the other parameters are the same as in Fig.~\ref{fig:nonMK-nonpos}.
    }
    \label{fig:anisostropy_NEP}
\end{figure}

\subsection{Non-energy preservation due to anisotropy}

The non-energy preserving condition $[H_k + H_j, U_{kj}] \neq 0$ can occur also when the exchange terms in the interaction Hamiltonian contains an anisotropic coupling. In such a case, the sector of $\Lambda_{n,n-1}$ operating on quantum coherences becomes fully coupled:
\begin{equation}
\label{eq}
\Lambda_{n,n-1} =
\begin{pmatrix}
a_n & 0 & 0 & b_n\\
0 & c_n & d_n & 0\\
0 & d_n^* & c_n^* & 0\\
1-a_n & 0 & 0 & 1-b_n
\end{pmatrix}.
\end{equation}
Hence, a condition for $d$ must be added to \eqref{eq:CP_conditions}, $|d_n|^2 \le b_n(1-a_n)$, and even in this more general case, the non-positivity of the KDQ distribution still probes $(a_n,b_n)$, albeit not $(c_n,d_n)$. Hence, \eqref{eq:KDvsCP} holds also in the non-energy preserving regime.

In Fig.~\ref{fig:anisostropy_NEP}, we report the behavior of $\mathcal{I}_{\mathrm{LFS}}$, $\mathcal{I}_{\mathrm{RHP}}$ and the cumulative non-positivity functional $\sum_n\mathcal{N}_q[P_n]$ as a function of the coupling anisotropy $\gamma$ for system-memory collisions. Specifically, keeping energy-preserving the interaction between memory and environment particles, we now model the collision between $S$ and $M$ with the anisotropic Hamiltonian, $H_{SM} {=} \frac{1{-}\gamma}{2}\sigma^x_S{\otimes}\sigma^x_M + \frac{1{+}\gamma}{2}\sigma^y_S{\otimes}\sigma^y_M + \sigma^z_S{\otimes}\sigma^z_M$.

All curves in the figure display a weak but systematic dependence on $\gamma$, with a local minimum around $\gamma{=}0$. The non-Markovianity measures remain within a narrow numerical range. This indicates that the impact of anisotropy on non-Markovianity is quite modest under energy-preserving memory-environment interactions.

\end{document}